\documentclass{article}
\usepackage{arxiv}

\usepackage[utf8]{inputenc} 
\usepackage[T1]{fontenc}    
\usepackage{hyperref}       
\usepackage{url}            
\usepackage{amsfonts}       
\usepackage{nicefrac}       
\usepackage{microtype}      
\usepackage{graphicx}
\usepackage{natbib}
\usepackage{doi}
\usepackage{multirow}
\usepackage{placeins}
\usepackage{siunitx} 
\usepackage{lmodern}
\usepackage{booktabs}

\def\topline{\toprule}
\def\midline{\midrule}
\def\botline{\bottomrule}

\makeatletter
\newcommand{\applabel}[2]{\def\@currentlabel{#1}\label{#2}}
\makeatother

\title{On the Predictive Skill of Artificial Intelligence-based Weather Models for \\ Extreme Events using Uncertainty Quantification}

\author{%
  Rodrigo Almeida \\
  Applied Machine Learning Group\\
  Fraunhofer Heinrich-Hertz Institute\\
  10587 Berlin, Germany \\
  \texttt{rodrigo.almeida@hhi.fraunhofer.de} \\
  \And
  Noelia Otero \\
  Applied Machine Learning Group\\
  Fraunhofer Heinrich-Hertz Institute\\
  10587 Berlin, Germany \\
  \AND
  Miguel-Ángel Fernández-Torres \\
  Signal Theory and Communications Dept.\\
  Universidad Carlos III de Madrid (UC3M)\\
  28911 Leganés, Madrid, Spain \\
  \And
  Jackie Ma \\
  Applied Machine Learning Group\\
  Fraunhofer Heinrich-Hertz Institute\\
  10587 Berlin, Germany \\
}

\begin{document}

\maketitle

\begin{abstract}
Accurate prediction of extreme weather events remains a major
challenge for artificial intelligence–based weather prediction systems. While deterministic
models such as FuXi, GraphCast, and SFNO have achieved competitive forecast skill
relative to numerical weather prediction, their ability to represent uncertainty
and capture extremes is still limited. This study investigates how state-of-the-art
deterministic artificial intelligence-based models respond to initial-condition perturbations
and evaluates the resulting ensembles in forecasting extremes. Using four
perturbation strategies (Gaussian, Perlin noise, Hemispheric Centered Bred
Vectors, and Huge Ensembles), we generate 50 member ensembles for the August 2022
Pakistan floods and China heatwave, and complement these case studies with a
global threshold-based evaluation. Ensemble skill is assessed against ERA5 and compared
with IFS ENS and the AIFS ENS probabilistic model using deterministic and probabilistic
metrics. Results show that simpler perturbations like Gaussian and Perlin noise
produce similarly realistic ensemble spread and probabilistic skill as flow-based approaches 
like HCBV and HENS, narrowing but not closing the performance gap with numerical weather 
prediction ensembles, or native probabilistic models which retain the highest probabilistic skill across variables. Model choice is the dominant factor
for ensemble performance, not perturbation method. Across variables, models capture temperature 
extremes more effectively than precipitation. These findings demonstrate that simple 
input perturbations can extend deterministic models toward 
probabilistic forecasting in hardware-constrained settings, 
supporting artificial intelligence-driven early warning systems.
\end{abstract}

\section{Introduction}

Artificial Intelligence (AI) is rapidly reshaping the field of weather and climate
forecasting, including the assessment and management of natural hazard risks
\citep{camps2025artificial}. Traditionally, Numerical Weather Prediction (NWP)
relies on solving complex systems of partial differential equations to simulate
atmospheric processes. However, this approach demands substantial computational
resources, which constrains both spatial and temporal resolution. Given the inherently
chaotic nature of the atmosphere, it is also crucial to account for
uncertainties in both the initial conditions and the model itself. Ensemble forecasting
addresses this challenge by generating multiple simulations to represent a range
of possible outcomes, forming the basis for probabilistic forecasts \citep{Leutbecher2008}.

With recent advances in machine learning (ML), the development of sophisticated
deep learning (DL) models, and the increasing availability of large datasets,
data-driven AI approaches are gaining traction, offering promising improvements
to traditional forecasting methods. AI-based weather prediction (AIWP)
models using transformers \citep{Bi2023,chen2023fengwupushingskillfulglobal, chenFuXiCascadeMachine2023, Allen2025},
Fourier neural operators \citep{bonevSphericalFourierNeural2023a, kurth2023},
and graph neural networks \citep{keisler2022,lamGraphCastLearningSkillful2023}
have pushed the skill of weather forecasting over what is considered the state of
the art for NWP models. Moreover, the recent operational integration of the AI
Forecasting System (AIFS), developed by the European Centre for Medium-Range
Weather Forecasts (ECMWF), alongside traditional NWP models, marks a
significant milestone in the evolution of AIWP \citep{langAIFSECMWFsDatadriven2024}.

Generative approaches have also been introduced, making use of diffusion-based
models such as GenCast \citep{priceGenCastDiffusionbasedEnsemble2024a},
which have the ability to directly produce a distribution of potential
future weather states. Similarly, \citet{Li2024} presented SEEDS, a Scalable
Ensemble Envelope Diffusion Sampler to generate an arbitrarily large
ensemble conditioned on as few as one or two forecasts from an operational
NWP system. Furthermore, hybrid approaches like NeuralGCM \citep{Kochkov2024}
combine the dynamical core of traditional NWP models with local ML-based parameterizations,
achieving performances comparable to operational ensemble forecasts.

Deterministic AIWP models now match or exceed deterministic NWP on standard
skill metrics \citep{raspWeatherBench2Benchmark2024a}, and have been shown
to capture and appropriately simulate physical laws in the atmosphere \citep{bano-medinaAreAIWeather2025}.
Their performance on extreme events, however, is more nuanced. Because they are
typically trained with a deterministic objective that rewards variance reduction,
they tend to smooth out the tails of the predictive distribution \citep{brenowitzPracticalProbabilisticBenchmark2024}.
\citet{olivettiDatadrivenModelsBeat2024} found that deterministic AIWP models
remain competitive with deterministic NWP for extremes, but their relative
skill depends strongly on variable, region, and lead time, performing best for
temperature, at short lead times, and near the tropics. \citet{zhangNumericalModelsOutperform2025}
and \citet{zhaoExtensionWeatherBench22025} report a similar variable-dependent
pattern, with stronger AIWP skill for warm extremes than for wet extremes,
and a widening gap relative to NWP at the most extreme tails. 

Deterministic skill alone, however, is insufficient for operational use.
Because the atmosphere is chaotic, any single forecast will diverge from reality
at longer lead times, and stakeholders need not just a best estimate but the
range of plausible outcomes and their probabilities \citep{gargWeatherBenchProbabilityBenchmark2022}.
This is especially true for extreme events, which are rare by definition and
frequently missed by deterministic forecasts at longer horizons.
Probabilistic forecasting, through ensembles, as exemplified by the ECMWF Integrated Forecasting System Ensemble (IFS ENS; \citet{molteni1996ecmwf}), hereafter referred to as ENS, is therefore widely recognized as the more
appropriate framework for high-impact situations, supporting decision-making
under uncertainty and enabling authorities to take preventive measures to protect
lives and infrastructure \citep{Ponzano2025}. 

Earlier work on AIWP was clearly centered on deterministic weather forecasting
due to the wide availability of data and benchmarks for such forecasts \citep{raspWeatherBench2Benchmark2024a}.
However, there has recently been a growing body of work focused on
probabilistic AIWP either by introducing post-hoc approaches akin to
uncertainty quantification \citep{bulteUncertaintyQuantificationDatadriven2024},
novel perturbation methods similar to those used in traditional NWP
ensembles \citep{maheshHugeEnsemblesPart2025}, by applying Monte Carlo
dropout \citep{gargWeatherBenchProbabilityBenchmark2022} or training the model
using a probabilistic objective through changes in the loss function, like
in AIFS ENS \citep{langAIFSCRPSEnsembleForecasting2024, zhongFuXiENSMachineLearning2025, aletSkillfulJointProbabilistic2025},
or by using generative diffusion AI models to directly produce probabilistic
outputs \citep{priceGenCastDiffusionbasedEnsemble2024a, couairon2024}.
Recent work suggests that the scarcity and imbalance of extreme events in
training datasets can limit model generalization on tail events \citep{sunPredictingTrainingData2025}.

Uncertainty quantification (UQ) can be done using post-hoc or following
approaches based on initial conditions. Post-hoc methods rely on an established
methodology to infer a probability distribution from a given deterministic model
output. An example of such methods is the EasyUQ method \citep{walzEasyUncertaintyQuantification2023}.
While post-hoc methods are attractive in theory since they are simpler to
implement, they make it harder to disentangle sources of uncertainty in the model.
Initial condition-based approaches involve making changes to the model input and producing 
an ensemble of forecasts to infer a probabilistic distribution. While previous studies 
have explored individual techniques such as Bred Vectors on single architectures 
\citep{bano-medinaCalibratedEnsemblesNeural2025, maheshHugeEnsemblesPart2025},
no prior work has compared how different AIWP architectures respond
to a common set of perturbation strategies, nor quantified the relative
contributions of model choice and perturbation choice to ensemble skill on
extremes.

This paper aims to drive the discussion on the usage of AIWP models, specifically
in the context of extreme weather forecasts, by comparing how different AIWP
architectures respond to different input perturbation methods. With this aim,
we hope to contribute to the following key research questions:
\begin{enumerate}
    \item How do perturbation-based ensembles built from deterministic AIWP models
        compare to natively probabilistic systems such as IFS ENS and AIFS ENS
        in capturing extremes, and how does this gap evolve with lead time
        and event rarity?

    \item Do AIWP ensembles capture different types of meteorological extremes
        with comparable skill, or are some variables systematically harder
        than others?

    \item To what extent is ensemble skill on extremes governed by the choice
        of model architecture versus the choice of perturbation method?
        
\end{enumerate}

By addressing these questions, our study seeks to advance the understanding of
both the potential and the limitations of deterministic data-driven models in
capturing extreme events through UQ. We assess these questions through a
comparison of multiple AIWP architectures under several perturbation
strategies, evaluated globally across multiple variables and percentile thresholds,
and illustrated in detail on two case studies: the 2022 Pakistan floods and the
August 2022 China heatwave. We anticipate that the insights gained in this
study will contribute to the practical deployment of data-driven models in
operational settings.

\section{Data}

Two data sources are considered in this study: 1) ERA5 reanalysis data \citep{hersbachERA5GlobalReanalysis2020, carverARCOERA5AnalysisReadyCloudOptimized2023},
which serves both as initial conditions for our forecasting models and as reference to evaluate the forecast performance; and 2) two ensemble forecasting
systems used as benchmarks: the ECMWF Integrated Forecasting System Ensemble
(IFS ENS) \citep{molteni1996ecmwf, ifsshort}, representing a state-of-the-art
numerical weather prediction ensemble, and the AIFS ENS \citep{langAIFSCRPSEnsembleForecasting2024},
representing a natively probabilistic AI-based ensemble. IFS ENS forecasts
were obtained from the WeatherBench 2 archive \citep{raspWeatherBench2Benchmark2024a}.
AIFS ENS forecasts were generated by the authors using the publicly released
model weights, initialized from ERA5 to ensure a consistent initialization across
all AIWP models evaluated in this study.  We use a subset of the ERA5
ARCO \citep{carverARCOERA5AnalysisReadyCloudOptimized2023} variables required
for prediction and evaluation, following \citet{bulteUncertaintyQuantificationDatadriven2024},
with 6-hourly temporal resolution (see Table \ref{era5} for the complete list).
To enable a fair comparison between AIWP models trained in the same dataset, all data
used in this study are downscaled to a common 1\textdegree × 1\textdegree ~spatial
resolution, using bilinear interpolation \citep{genevaNVIDIAEarth2Studio2024}.
The evaluation is conducted on: Geopotential at 500 hPa ($Z500$), Temperature
at 850 hPa ($T850$), 2 m Temperature ($T2M$), U and V components of wind at 10 m ($U10M$, $V10M$), Total
Precipitation (6h accumulation, $TP_{6h}$), and additionally 24h aggregations
for total precipitation ($TP_{24h}$) and maximum for all other evaluation variables 
($T2M_{24h}$, $T850_{24h}$, $Z500_{24h}$, $U10M_{24h}$, $V10M_{24h}$).

\begin{table}[h]
\caption{ERA5 variables used for forecasting with AIWP models.}\label{era5}
\begin{center}
\begin{tabular}{lp{0.37\textwidth}ll}
\topline
\textbf{Variable} & \textbf{Description} & \textbf{Unit} & \textbf{Pressure levels (\si{hPa})} \\
\midline
\textbf{z (levels)}\textsuperscript{\textit{a}}
  & Geopotential 
  & \si{m^{2}.s^{-2}} 
  & \multirow{7}{*}{\begin{tabular}[c]{@{}l@{}}1000, 925, 850, 700, \\ 600, 500, 400, 300, \\ 250, 200, 150, 100, \\ 50 \quad (13 levels)\end{tabular}} \\
\textbf{q} & Specific humidity & \si{kg.kg^{-1}} & \\
\textbf{r}\textsuperscript{\textit{b}} & Relative humidity & \% & \\
\textbf{t}\textsuperscript{\textit{a}} & Temperature & \si{K} & \\
\textbf{u}\textsuperscript{\textit{a}} & U component of wind & \si{m.s^{-1}} & \\
\textbf{v}\textsuperscript{\textit{a}} & V component of wind & \si{m.s^{-1}} & \\
\textbf{w} & Vertical velocity & \si{Pa.s^{-1}} & \\
\midline
\textbf{msl}\textsuperscript{\textit{a}} & Mean sea level pressure & \si{Pa}  \\
\textbf{u10m}\textsuperscript{\textit{a}} & 10m U component of wind & \si{m.s^{-1}}  \\
\textbf{v10m}\textsuperscript{\textit{a}} & 10m V component of wind & \si{m.s^{-1}}  \\
\textbf{u100m} & 100m U component of wind & \si{m.s^{-1}}  \\
\textbf{v100m} & 100m V component of wind & \si{m.s^{-1}}  \\
\textbf{t2m}\textsuperscript{\textit{a}} & 2m Temperature & \si{K}  \\
\textbf{d2m} & 2m Dewpoint temperature & \si{K}  \\
\textbf{stl1} & Soil temperature level 1 & \si{K}  \\
\textbf{stl2} & Soil temperature level 2 & \si{K}  \\
\textbf{sp} & Surface pressure & \si{Pa}  \\
\textbf{tcwv} & Total column water vapour & \si{kg.m^{-2}}  \\
\textbf{tcw} & Total column water & \si{kg.m^{-2}}  \\
\textbf{swvl1} & Volumetric soil water layer 1 & \si{m^{3}.m^{-3}} \\
\textbf{swvl2} & Volumetric soil water layer 2 & \si{m^{3}.m^{-3}} \\
\textbf{tp06}\textsuperscript{\textit{c}} & Total precipitation (6h accumulation) & \si{m}  \\
\textbf{z (surface)} & Geopotential at surface & \si{m^{2}.s^{-2}}  \\
\textbf{lsm} & Land--sea mask & (0--1)  \\
\textbf{sdor} & Standard deviation of orography & \si{m} \\
\textbf{slor} & Slope of sub gridscale orography & \si{m} \\
\botline
\end{tabular}
\end{center}
\footnotesize{\textsuperscript{\textit{a}} This variable is perturbed when using perturbation methods.}
\footnotesize{\textsuperscript{\textit{b}} This variable is derived from pressure, \textbf{q}, and \textbf{t}.}
\footnotesize{\textsuperscript{\textit{c}} This variable is accumulated in 6-hourly intervals from the original 1-hourly data.}
\end{table}

\section{Methodology}

\begin{figure}[h]
    \centerline{\includegraphics{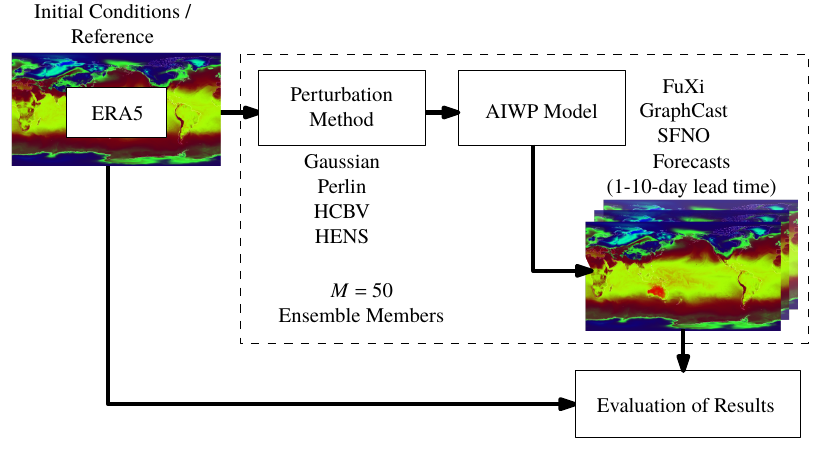}}
    \caption{Overview of the study methodology, including the initial condition
    perturbation, forecasting with AIWP models, and evaluation of results stages.}
    \label{methodology}
\end{figure}

Figure \ref{methodology} illustrates the methodology followed in this study.
First, starting from ERA5 reanalysis data as initial conditions, we apply
several perturbation methods, each of them generating $M=50$ different initial
states. Then, these states are provided to diverse 6-hourly resolution AIWP models,
resulting in $M=50$ different predictions, which constitute ensemble forecasting
members for up to a 10-day lead time. The ensembles for all perturbation
methods and AIWP models, along with ENS and AIFS ENS, are finally evaluated 6-hourly and
at daily resolution against the reference ERA5 to assess their forecasting capabilities.

For data handling and running predictions with various models, we use the
NVIDIA-developed Earth2Studio framework \citep{genevaNVIDIAEarth2Studio2024}.
This modular and Python-based framework enables running large-scale
inference, with multiple models and perturbation methods. In addition, for the
evaluation of the forecasts, we make use of the WeatherBench framework
\citep{raspWeatherBench2Benchmark2024a} and the probabilistic metrics extension
to WeatherBench introduced in \citet{zhaoExtensionWeatherBench22025}.

\subsection{Models}

We select three AIWP models for the purpose of this study that fulfill the following
conditions:

\begin{enumerate}
    \item Perform at/close to the state of the art for deterministic metrics
        and medium-range weather forecasting;

    \item Have both model code implementation and weights available in an open source way;

    \item Produce forecasts for total precipitation and surface temperature,
        given that we want to understand the different models' prediction behavior
        concerning extreme precipitation and heatwaves, respectively.
\end{enumerate}

Based on these criteria and our prior literature review, as well as taking into
account diversity in model architecture type, this study uses three
different architectures: FuXi \citep{chenFuXiCascadeMachine2023}, GraphCast \citep{lamGraphCastLearningSkillful2023},
and SFNO (Spherical Fourier Neural Operator) \citep{bonevSphericalFourierNeural2023a}.
For comparison, we also evaluated AIFS ENS \citep{langAIFSCRPSEnsembleForecasting2024},
a probabilistically trained AIWP example model. Below, we summarize these
models to highlight their architectures and forecasting approaches, referring
the reader to their original publications for further details.

\textbf{FuXi} is a hybrid architecture combining a cascading U-Transformer
with Swin Transformer V2 blocks to deliver high-accuracy forecasts across
multiple time scales. FuXi uses three distinct sets of model weights to predict
the weather between 0-5 days (FuXi short), 5-10 days (FuXi medium), and over
10 days (FuXi long) \citep{chenFuXiCascadeMachine2023}. As input, it takes
two timesteps to predict the subsequent meteorological state (
$X^{t-1}, X^{t}\rightarrow X^{t+1}$ ).

\textbf{GraphCast} builds on the pioneering work of \citet{keisler2022}
using a GNN based on an encoder, processor, and decoder architectures that
include simultaneous multi-mesh message passing at different scales \citep{lamGraphCastLearningSkillful2023}.
It relies on two timesteps as input to predict the following timestep (
$X^{t-1}, X^{t}\rightarrow X^{t+1}$ ), along with multiple forcing fields like
time of day, day of year, and the total top of atmosphere incident solar radiation,
among others. GraphCast has been trained on ERA5 data at 1\textdegree~resolution
and fine-tuned for IFS initial conditions at IFS-like resolutions (0.25\textdegree).
As stated in the previous section, we decided to use the 1\textdegree~model
for GraphCast in order to enable a fair comparison with all other models, which
have not been finetuned on IFS initial conditions. 

\textbf{SFNO} is a Spherical Fourier Neural Operator-based architecture that
is designed to leverage the properties of the Fourier operators on spherical
geometries, modeling long-range dependencies in spatio-temporal data via convolutions
\citep{bonevSphericalFourierNeural2023a}. This model does not predict total precipitation
directly, but comes with a trained diagnostic model \citep{nvidiangccatalogPhysicsNeMoCheckpointsAFNO_DX_TPV1ERA52025}
that takes in the output of the SFNO model and produces a total precipitation
output ( $X^{t}\rightarrow X^{t+1}, \omega(X^{t+1})$, where $\omega$ is the
diagnostic total precipitation model).

While this study focuses on three deterministic, state-of-the-art AIWP
models, we acknowledge that other architectures could also have been included.
In particular, models such as AIFS \citep{langAIFSECMWFsDatadriven2024} and
NeuralGCM \citep{Kochkov2024} represent important advances and improvements in
forecast skill. Nevertheless, we believe that the selected set provides a representative
cross-section of current state-of-the-art approaches across transformer-, GNN-,
and operator-based architectures, allowing for a comparison of how leading
AIWP models respond to input perturbations.

\subsection{Perturbation Methods}
\label{perturbation}

We select four perturbation methods for this study, spanning a range of
complexity from simple stochastic baselines to flow-dependent approaches. The
first two methods, Gaussian noise and Perlin noise, generate spatially correlated
perturbations that are independent of the atmospheric state, differing in
how spatial correlation is introduced: through a prescribed covariance function
(Gaussian) or through procedural interpolation on a lattice (Perlin). The
remaining two methods, the Hemispheric Centered Bred Vector (HCBV) and Huge Ensembles
(HENS), are both based on bred vectors, a class of flow-dependent, nonlinear
perturbations designed to capture the fastest-growing modes of the
atmospheric flow \citep{fengComparisonNonlinearLocal2018, tothEnsembleForecastingNCEP1997a}.
HCBV follows the formulation of \citet{bano-medinaCalibratedEnsemblesNeural2025},
while HENS incorporates additional model-specific amplitude scaling as
introduced by \citet{maheshHugeEnsemblesPart2025}. This selection allows us
to assess both whether spatial correlation configuration in the perturbations matters (Gaussian
vs.\ Perlin) and whether flow-dependent structure provides additional skill beyond
appropriately scaled stochastic noise (Gaussian/Perlin vs.\ HCBV/HENS).

For all perturbation methods, only the input variables common to all three AIWP
models are perturbed (see Table~\ref{era5}). Each method relies on a
dimensionless scaling factor $s$ that controls the overall perturbation
magnitude. To select an appropriate value, we performed a sensitivity
analysis over $s$ for each perturbation method (see Appendix~\ref{scaling}). Based on this
analysis, we set $s = 0.35$ for Gaussian and HENS perturbations. These scaling
factors are consistent with the scaling used in \citet{maheshHugeEnsemblesPart2025}.
For HCBV, we found the optimal scale factor as one order of magnitude
smaller ($s = 0.035$), while for Perlin noise the optimal factor is one-quarter
of the $0.35$ ($s = 0.0875$). Example input perturbations for each of the methods described are available in Appendix~\ref{perturbations}.

\subsubsection{Gaussian Noise}
The Gaussian perturbation method generates ensemble members by adding a realization
of a Gaussian random field on the sphere to the initial conditions. This
field is sampled spectrally using spherical harmonics, with a Matern covariance
function that controls the spatial correlation structure. The perturbed
initial condition at each grid point is given by:
\begin{equation}
    y_{\text{pert}} = y + s \cdot \sigma_{\text{clim}} \cdot \xi
\end{equation}
\noindent
where $y$ is the unperturbed initial condition, $s$ is a dimensionless
scale factor, $\sigma_{\text{clim}}$ is the climatological standard deviation
of the variable (computed from the ERA5 1990--2020 climatology),
and $\xi$ is a mean-zero, unit-variance spherical  Gaussian random field evaluated at each latitude and longitude. 
The Matern covariance uses the correlation length scale of 3 and smoothness of 4
set by the default configuration of the \texttt{SphericalGaussian}
implementation in Earth2Studio \citep{genevaNVIDIAEarth2Studio2024}. 

\subsubsection{Perlin Noise}

Perlin noise is a procedural noise function that generates spatially correlated
random fields by interpolating between pseudo-random gradient vectors on a regular
lattice, following the configuration in \citet{Bi2023}. Unlike the Gaussian
method, which derives its correlation structure from a prescribed covariance
function, Perlin noise produces spatial coherence through local
interpolation. The perturbed initial condition is given by:
\begin{equation}
    y_{\text{pert}} = y + s \cdot \sigma_{\text{clim}} \cdot P
\end{equation}
\noindent
where $P \in [-1, 1]$ is the Perlin noise value at each
latitude and longitude, and all other terms are as defined
above. Each ensemble member uses an independent random seed, producing a unique
perturbation field.

\subsubsection{Hemispheric Centered Bred Vector (HCBV)}
The bred vector algorithm, introduced by \citet{Toth1993}, estimates the
fastest-growing perturbation modes by exploiting the fact that initial
conditions produced by data assimilation accumulate growing errors. Starting
from a small seed perturbation, the method uses the forecast model itself to
identify the directions in the state space along which errors amplify fastest.

A seed perturbation $\Delta z_{500}$ is generated by adding a correlated 
spherical Gaussian noise field (with $\sigma _{\text{clim}}$ amplitude) to 
the 500\,hPa geopotential variable. The AIWP model
is then run from both the perturbed and unperturbed initial conditions to produce
forecasts $f_{p}$ and $f_{u}$, respectively. The raw bred vector is computed
as the difference between these two forecasts, normalized separately over
each hemisphere and scaled:
\begin{equation}
    \Delta f = s \cdot \sigma_{\text{clim}}\cdot \frac{f_{p}- f_{u}}{h(f_{p}-
    f_{u})}
\end{equation}
\noindent
where $s$ is the dimensionless scale factor, $\sigma_{\text{clim}}$ is the
climatological standard deviation of the variable (computed from the ERA5 1990--2020
climatology), and $h(\cdot)$ denotes the hemispheric norm, computed separately
for the North and South regions (defined as
$|\text{latitude}| > 70^{\circ}$) and interpolated in the tropics, following \citet{maheshHugeEnsemblesPart2025}. 

To better capture the dominant growing modes, this breeding cycle is
repeated $d = 3$ times: at each cycle, the bred vector from the previous
iteration is used as the new seed perturbation, and the model is run forward
for one autoregressive step (6\,h) to recompute the normalized difference.
The three recursive cycles span a total breeding period of 18\,h, yielding
the final perturbation $\Delta f_{d}$. This integration depth follows \citet{bano-medinaCalibratedEnsemblesNeural2025}
and is intended to allow the perturbation to project more strongly onto the dynamically
unstable modes of the atmosphere.

\begin{figure}[h]
    \centerline{\includegraphics{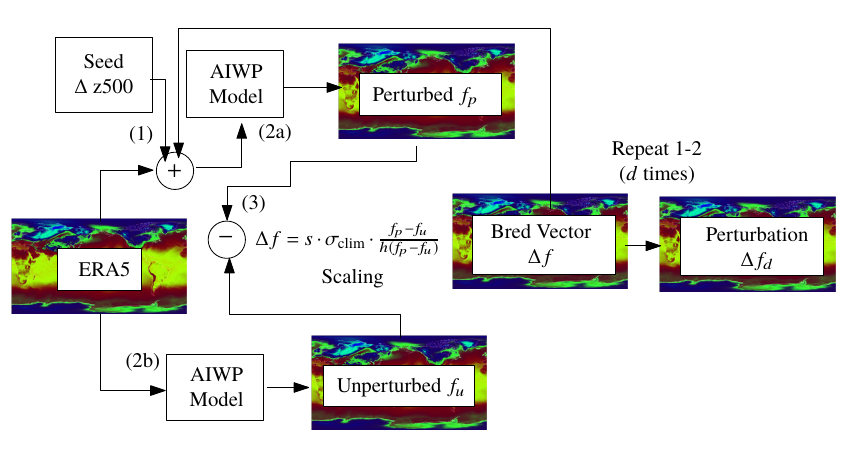}}
    \caption{Hemispheric Centered Bred Vector (HCBV) perturbation method. A seed
    perturbation $\Delta z_{500}$, generated as a correlated spherical
    Gaussian noise field, is added to the 500\,hPa geopotential variable.
    The AIWP model produces perturbed ($f_{p}$) and unperturbed ($f_{u}$)
    forecasts, whose difference is normalized separately for the North and
    South regions ($|\text{latitude}| > 70^{\circ}$) with interpolation in the
    tropics, denoted by $h(\cdot)$. The result is scaled by the dimensionless
    factor $s$ and the climatological standard deviation $\sigma_{\text{clim}}$.
    This breeding cycle is repeated $d = 3$ times, each corresponding to one
    6\,h autoregressive model step (18\,h total), yielding
    the final perturbation $\Delta f_{d}$. The perturbation is additionally
    centered: $\Delta f_{d}$ is alternately added and subtracted from the initial
    conditions. Diagram adapted from
    \protect\citet{bano-medinaCalibratedEnsemblesNeural2025}.}
    \label{hcbv}
\end{figure}

The perturbation is additionally centered: for each pair of ensemble members,
the bred vector $\Delta f_{d}$ is alternately added to and subtracted from the
initial conditions, ensuring that the ensemble mean remains unbiased with
respect to the unperturbed analysis. The method is illustrated in Fig.~\ref{hcbv}.

\subsubsection{Huge Ensembles (HENS)}

Huge Ensembles (HENS) is an adapted version of the HCBV method, introduced by
\citet{maheshHugeEnsemblesPart2025}, that incorporates model-specific amplitude
tuning to better match the expected forecast error growth of each AIWP architecture.
The key modification is in the scaling of the perturbation: HCBV uses a climatological
scaling, HENS replaces this with a model dependent amplitude:
\begin{equation}
    \Delta f = s \cdot \text{RMSE}_{48h}\cdot \frac{f_{p}- f_{u}}{h(f_{p}- f_{u})}
\end{equation}
\noindent
where $\text{RMSE}_{48h}$ is the global root mean squared error of the deterministic
AIWP model at a 48\,h lead time for the perturbed variable, and $s$ is the
dimensionless scale factor. This scales the perturbation to match with the actual
forecast error magnitude of each model, so that models with larger intrinsic
errors receive proportionally larger perturbations. The same model-specific scaling
is also applied to the seed perturbation $\Delta z_{500}$.

It should be noted that the original HENS implementation \citep{maheshHugeEnsemblesPart2025}
also employs an ensemble of model checkpoints to produce the final ensemble
prediction, combining both initial-condition and model-induced diversity. In
this study, we isolate the HENS initial-condition perturbation method only,
in order to enable a direct comparison between perturbation strategies.

\subsection{Evaluation Metrics}

We select three complementary evaluation metrics to assess both deterministic
and probabilistic forecast skill.

The \textbf{Root Mean Squared Error (RMSE)} measures the average magnitude
of errors between the ensemble mean forecast or deterministic forecast and
the reference values, penalizing larger errors more heavily. Lower RMSE
indicates better deterministic accuracy. An RMSE of zero corresponds to a perfect
forecast. 

The \textbf{Continuous Ranked Probability Score (CRPS)} evaluates the full
predictive distribution by measuring the integrated squared difference
between the forecast cumulative distribution function and the reference. It
generalizes the mean absolute error to probabilistic forecasts,
simultaneously rewarding both narrow predictive distributions
and statistical consistency with the reference. Lower CRPS indicates
better probabilistic skill, with zero representing a perfect probabilistic
forecast.

The \textbf{Receiver Operating Characteristic Skill Score (ROCSS)} quantifies
the ability of a forecast to discriminate between the occurrence and non-occurrence
of a binary event, relative to random chance. It is derived from the Area Under
the ROC Curve (AUC), which measures the trade-off between hit rate and false
alarm rate across all probability thresholds. A ROCSS of 1 indicates perfect
discrimination, 0 indicates no skill, and negative values indicate performance
worse than random chance. This metric becomes relevant for extremes through
the choice of an event threshold $\tau$. In this study, we define events
using climatological percentiles of the ERA5 1990--2020 distribution (primarily
$\tau = P_{99}$, with additional analysis at the 90th, 95th, 98th, and 99.5th
percentiles), so that the ROCSS specifically measures discrimination skill for
rare, high-impact events. Throughout this study we made use of daily aggregates
(sum for precipitation, maxima for all other variables) for computing
ROCSS following \citet{zhaoExtensionWeatherBench22025}. 

To quantify the relative importance of model architecture versus
perturbation method in determining ensemble skill, we apply a two-way analysis
of variance (ANOVA) to each forecast metric and report the variance fraction
$\eta^{2}$ attributable to model architecture, perturbation method, and their
interaction. Variance-partitioning approaches of this kind have
a precedent in the climate-projection literature, where ANOVA frameworks
are used to decompose total ensemble variance into contributions from modeling
factors and their interactions \citep{yipSimpleCoherentFramework2011}.

Formal definitions of all three metrics and the ANOVA approach are provided in Appendix~\ref{metricsapp}.

\section{Case studies}
\label{casestudy}

\begin{figure}[ht]
    \centerline{\resizebox{\textwidth}{!}{\includegraphics{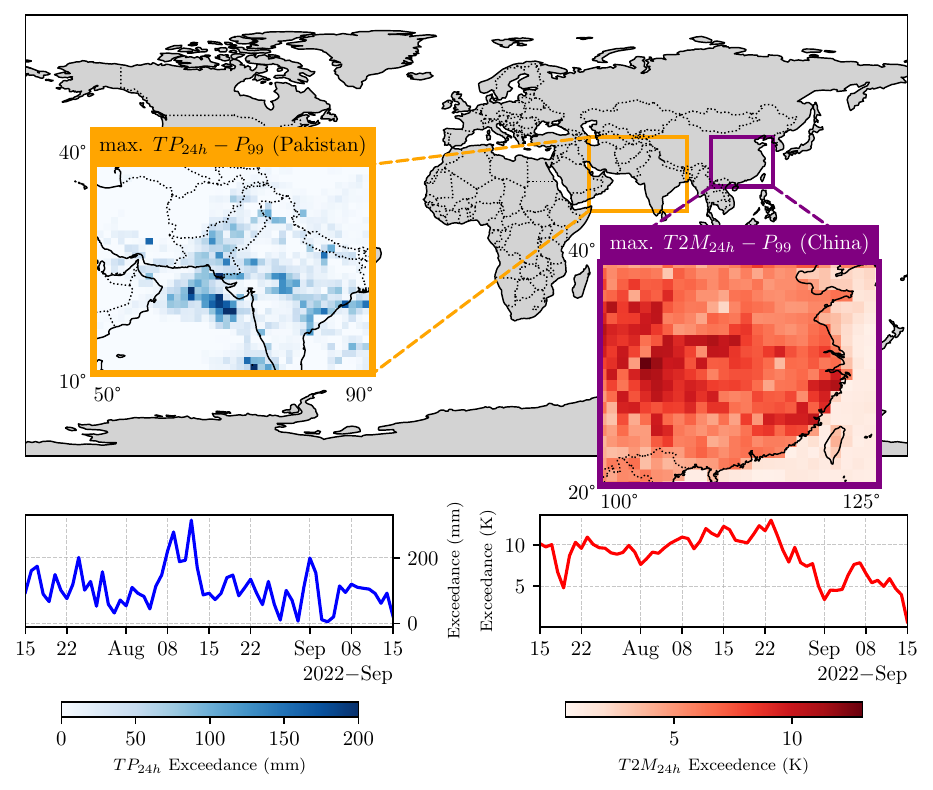}}}
    \caption{Analysis of the August 2022 Pakistan extreme precipitation and
    China heatwave, based on ERA5. The values displayed (in the spatial and temporal
    dimensions) correspond to the exceedance of the 99th percentile of the
    ERA5 1990-2020 climatology for daily total precipitation and daily
    maximum temperature. The geographical bounds of each case study are also
    shown. A maximum of 200 mm of precipitation exceedance can be
    appreciated in Pakistan, while a sustained 10 K exceedance over August is
    observed in China, further reinforcing the extreme nature of both events. }
    \label{casestudiesmap}
\end{figure}

Two case studies are considered: the Pakistan Floods and the China
Heatwave in August 2022. They are illustrated in Figure \ref{casestudiesmap}.
For our purposes, only data within the geographical bounds displayed in the figure
and corresponding to August 2022 is used. For both the Pakistan floods and
the China heatwave, we use the climatological 99th-percentile threshold to isolate
and define the extremes, consistent with precipitation analyses in \citet{Singh2024-wk}
and heatwave characterizations in \citet{sunHeatWaveCharacteristics2021}.

\subsection{Summer 2022 Pakistan Floods}

In the summer of 2022, Pakistan experienced catastrophic flooding that submerged
one-third of the country, resulting in over 1000 deaths, displacing 30 million
people, and causing over 30 billion USD in damages \citep{Arshad2025-wd}.
The rainfall observed in this period was four standard deviations above the
climatological mean and twice the amount of a previous flooding event in
2010 \citep{hongCauses2022Pakistan2023}. Recent studies have linked this
extreme event with concurrent heatwaves \citep{hongCauses2022Pakistan2023},
which favored snow melt, subsequently followed by extreme precipitation \citep{Shehzad2023}.
During that time, parts of the territory experienced daily precipitation
amounts above 100 mm, which, according to \citet{ikramFutureTrendsFrequency2016},
is considered extreme in Pakistan. Large parts of Pakistan with daily precipitation
exceeding 100 mm during August 2022 can be observed in Figure~\ref{casestudiesmap} 
corresponding to the 99th percentile of the 1990-2020 climatology of daily precipitation.

\subsection{August 2022 China Heatwave}

During August 2022, southern China was hit with a record-breaking heatwave, with
over 360M people experiencing temperatures above $40^{\circ}$C \citep{gongAttributionAugust20222024}.
This heatwave represented the greatest anomaly since 1979 and caused a severe impact,
including massive droughts \citep{zhouExtremeHeatWave2023, SunBongCharacteristicsCausesHotdry2023}.
As can be seen in Figure~\ref{casestudiesmap}, the region presented close to
$10^{\circ}$C exceedance with respect to the 99th-percentile of the ERA5 1990-2020
climatology for the daily maximum temperature on the majority of days in August 2022,
with an even greater exceedance in the Sichuan Basin.

\section{Results}
We begin this section by evaluating the performance of the three selected
deterministic AIWP models in predicting extreme events under input
perturbations. The resulting ensembles exhibited varying skill depending on the
predicted variable, model architecture, perturbation strategy, and forecast
lead time. In the following figures, we analyze the ROCSS for 90th to 99.5th percentile
thresholds. Then, for each case study, we analyze the ROCSS skill and inspect
the spatial spread characteristics. Finally, we evaluate the relative
contribution of model and perturbation choice in the forecast skill. In addition,
we explore the links between the performance in these case studies and
global evaluation metrics like CRPS and RMSE. 

\begin{figure}[h]
    \centerline{ \resizebox{\textwidth}{!}{\includegraphics{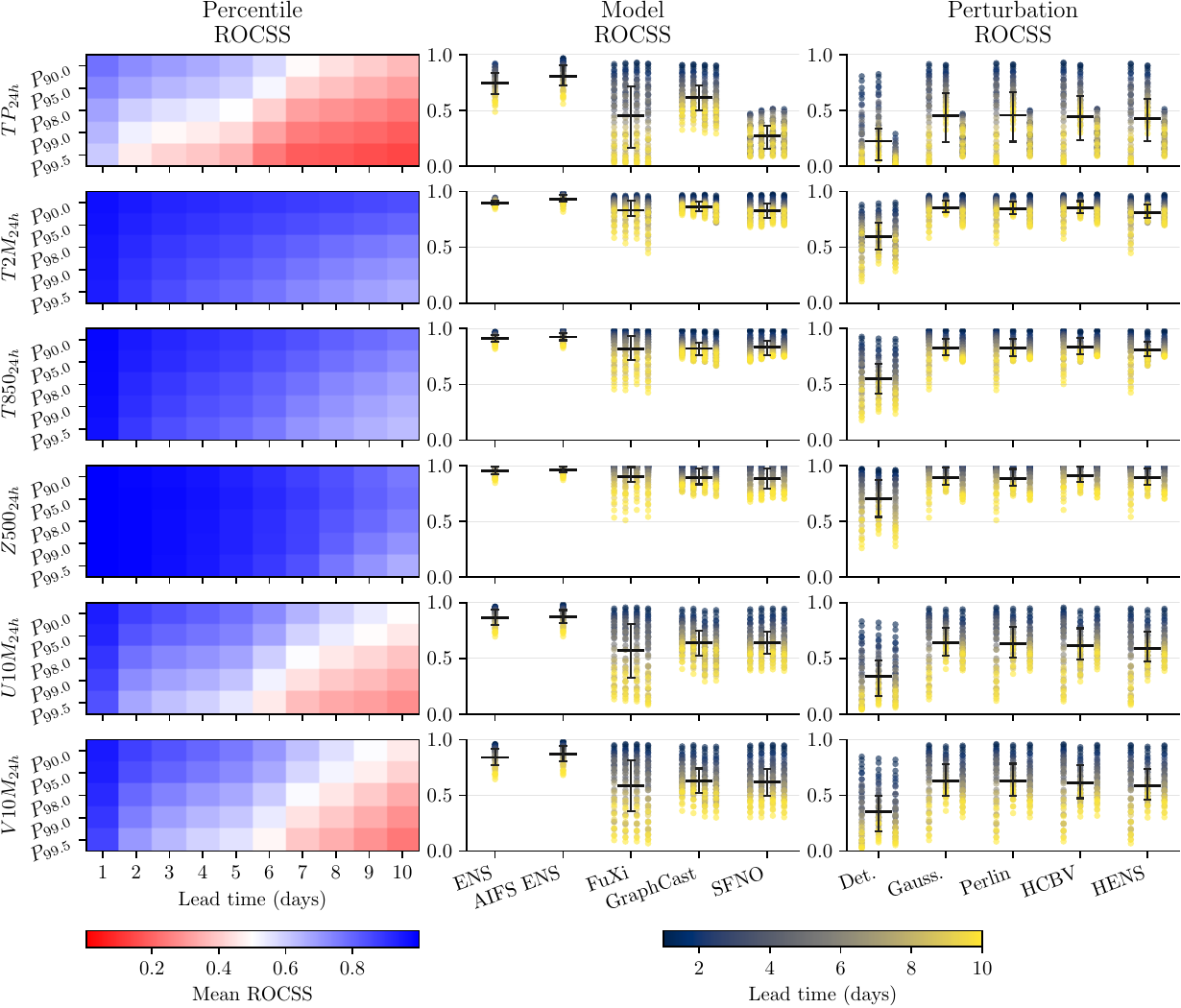}}}
    \caption{Global ROCSS decomposed by percentile threshold, model architecture,
    and perturbation method across August 2022. In the left column, we display mean ROCSS for models and perturbations as a function of lead time and percentile threshold; blue indicates high skill, red low skill. In the middle and right columns, ROCSS per model and ROCSS per perturbation, with AIFS~ENS and ENS as benchmarks, and deterministic forecasts. Each dot in this plot represents the metric for a given lead time, percentile, model, and perturbation, with each column grouping dots per perturbation (Gaussian, Perlin, HCBV, and HENS) or model (FuXi, GraphCast, SFNO). The similar spread of ROCSS in the right column indicates that model architecture is the dominant factor governing discrimination skill, while the choice of perturbation method has a comparatively small effect.}
    \label{summaryglobalrocss}
\end{figure}

Globally, in Fig.~\ref{summaryglobalrocss}, we can conclude that discrimination
skill decreases as percentile increases across models and variables (i.e., forecast
skill is reduced in more extreme events). Overall, forecast skill for total precipitation
is the lowest, followed by 10m wind vectors. Geopotential shows the best discrimination
skill. Looking at the model, we can see stark performance differences across
models for precipitation. Performance decreases with lead
time, with FuXi displaying the highest spread across lead times. ENS and AIFS
ENS outperform on average for discrimination skill. Analyzing
perturbations, we observe that the shape of the spread across perturbation methods is very similar, even in precipitation, where the most pronounced model differences are found. This indicates a comparatively small effect of perturbation choice on forecast skill.

\begin{figure}[h]
    \centerline{ \resizebox{\textwidth}{!}{\includegraphics{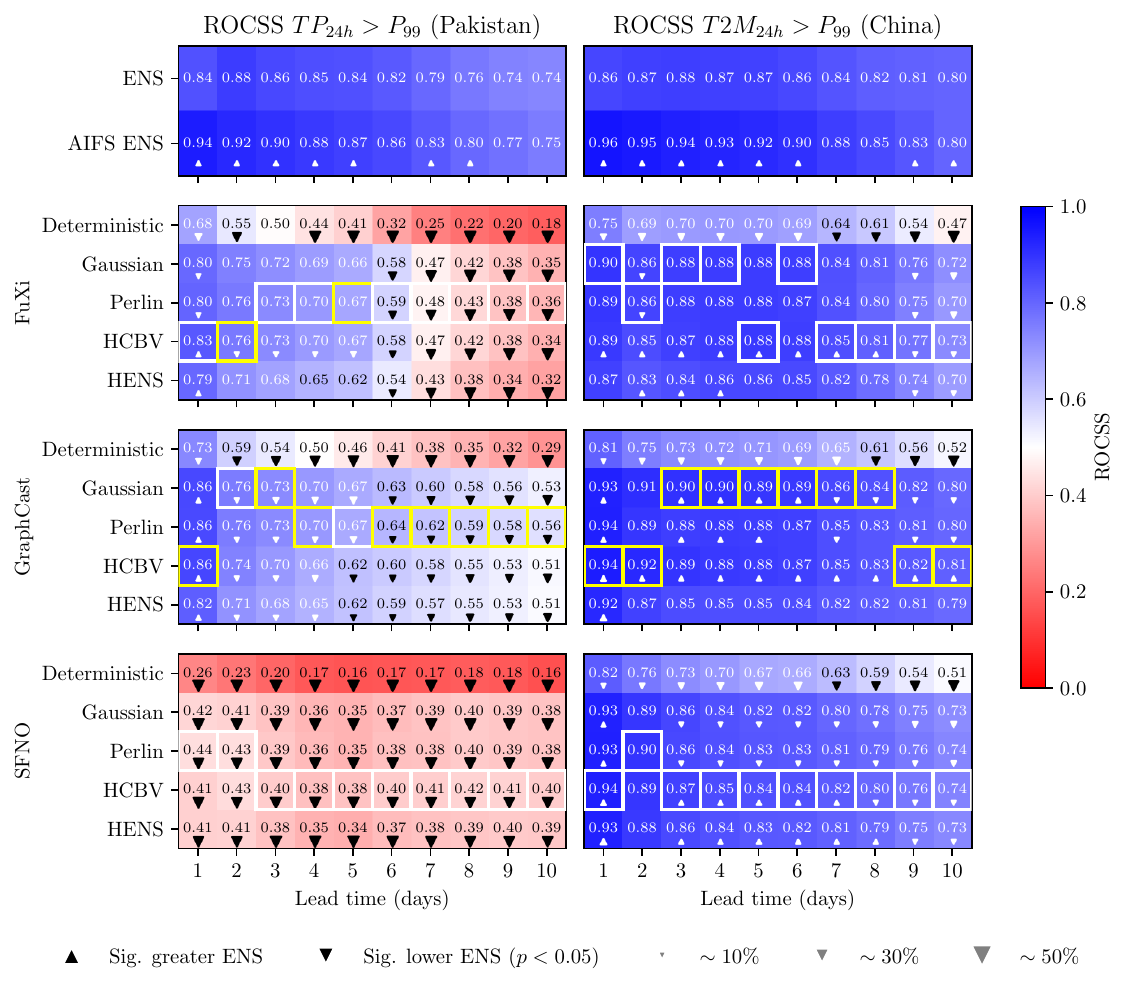}} }
    \caption{ROCSS at the 99th percentile for daily precipitation (Pakistan,
    left) and maximum daily temperature (China, right) in August 2022. ENS
    and AIFS~ENS benchmarks are shown at the top, followed by each AIWP model
    with its deterministic forecast and four perturbation methods. Triangles
    indicate statistically significant differences relative to ENS ($p < 0.05$):
    upward for higher, downward for lower, with size proportional to the magnitude
    of the difference. Yellow boxes represent the highest metric overall, and white
    boxes highlight the highest metric for each model. All perturbation methods
    improve upon the deterministic baseline, but differences between model architectures
    are substantially larger than differences between perturbation strategies.}
    \label{rocss}
\end{figure}

Figure~\ref{rocss} displays our case study evaluation, with the ROCSS at the
99th-percentile for daily precipitation over Pakistan and daily maximum temperature
over China. Overall, ensemble skill decreases with lead time, with the
exception of SFNO for precipitation, reflecting the limited predictability
horizon of these AIWP and NWP systems. For both ENS and AIFS ENS, we notice
a slight increase in performance for precipitation from 1 to 2 lead times, most
likely because ENS is initialized with IFS initial conditions and
AIFS ENS is also fine-tuned with that data, which causes a small dip in
performance in the first forecast step when compared with models trained
with ERA5. However, both ENS and AIFS ENS showcase the best metrics across
variables and lead times, with AIFS ENS showing significantly higher ROCSS than
ENS up to a 5-day lead time for precipitation and 6-day lead time for temperature,
demonstrating the benefits of a probabilistically trained model. Among the perturbation
methods, Gaussian, Perlin, and HCBV achieved similarly high and the highest probabilistic
skill on these extreme events, and all perturbation methods improved upon
the model's deterministic skill.

For very high precipitation (i.e., 99th percentile) in Pakistan, GraphCast–HCBV
(and Gaussian and Perlin) reaches ROCSS $\approx 0.86$ at lead time of 1 day,
the highest ROCSS for this variable. At longer lead times, GraphCast–Perlin has
the best discrimination skill, keeping ROCSS above $0.6$ until 7 days
lead time. In contrast, SFNO ensembles exhibit weak discrimination skill for
extreme precipitation (ROCSS $< 0.44$), indicating very limited performance for
precipitation extremes. Interestingly, the performance for SFNO overall decreases
around a 4-5-day lead time and recovers around the 8-day lead time for HCBV perturbations.
SFNO uses a separate diagnostic model to produce precipitation forecasts. This
diagnostic model was trained using ERA5 data \citep{nvidiangccatalogPhysicsNeMoCheckpointsAFNO_DX_TPV1ERA52025},
not SFNO forecasts, which could potentially explain the poor performance of
SFNO for precipitation overall.

For temperature extremes (i.e., 99th percentile) in China, GraphCast–Gaussian
performs best, attaining ROCSS $\approx 0.94$ at a lead time of 1 day, nearly
matching AIFS ENS ($0.96$), and being significantly higher than ENS up to a
6-day lead time. FuXi–HENS shows the worst performance, and it is the only
model that does not show significantly higher ROCSS than ENS at any lead
time. For precipitation, Perlin and Gaussian perturbations support
GraphCast ROCSS for longer lead times than FuXi, which very precipitously decreases performance across lead times, from $0.83$ to
$0.32$. In addition, for all models, ROCSS for temperature is comparatively much
more stable across lead times. This pattern suggests that AIWP models are
more effective at representing smoother, large-scale thermodynamic
fields than precipitation fields, and perturbations alone are not able to overcome that for discrimination skill, although they do seem to support GraphCast's performance more than FuXi or SFNO.

\begin{figure}[t]
    \centerline{ \includegraphics[width=0.8\textwidth]{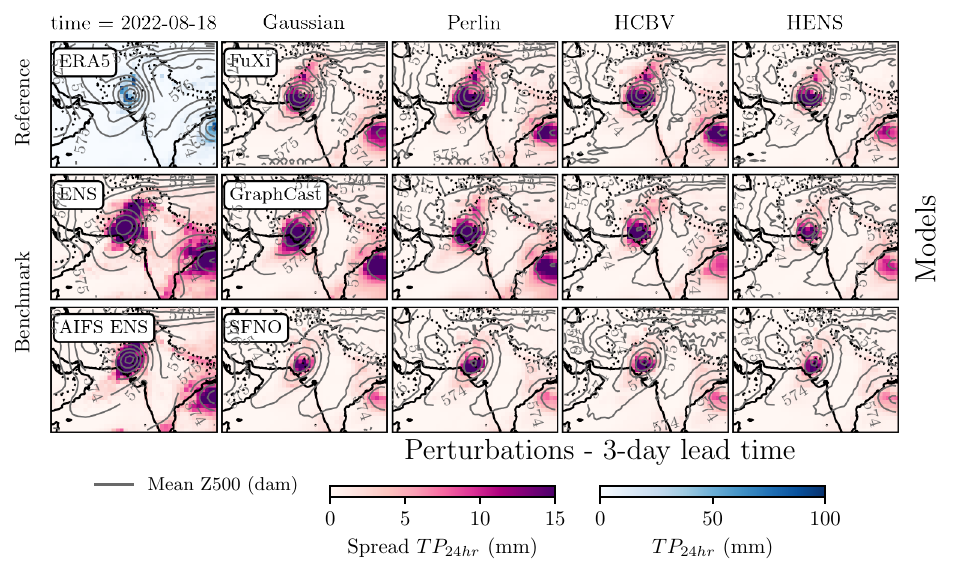} }
    \caption{Daily accumulated precipitation spread for the different
    ensemble models (ENS, AIFS ENS, AIWPs) over Pakistan on 18th August 2022
    for a 3-day lead time forecast. ERA5 spatial distribution for
    daily precipitation is shown in the top left corner. Daily average
    geopotential height at 500 hPa is also displayed as contour lines. FuXi and
    GraphCast with Gaussian/Perlin noise most closely match the ENS spread.
    Within each model, the perturbations look very similar.}
    \label{spreadpakistan}
\end{figure}

The spatial ensemble spread at 3-day lead time in Pakistan (Fig.~\ref{spreadpakistan})
and China (Fig.~\ref{spreadchina}) highlights how model selection plays the
biggest role in the ensemble spread. In Fig.~\ref{spreadpakistan}, we can observe
that ENS and AIFS ENS capture the heavy precipitation core and associated uncertainty.
The perturbed AIWP ensembles show similar spatial spread composition across
perturbations, although only FuXi and GraphCast with Gaussian and Perlin seem
to be able to reproduce the uncertainty over Pakistan and in eastern India. The perturbations are however slightly underdispersive when compared with the benchmark. Similar patterns are observed for temperature.

We further assess global performance to determine whether the patterns
observed in these case studies extend across all regions and non-extreme settings.
Globally averaged metrics reinforce the regional findings (see Appendix~\ref{globalplots}). All
perturbations yield similar probabilistic metrics for each model, although
Gaussian and Perlin have a slight edge (see Fig. \ref{crpsplot}). Probabilistic skill is consistently higher for temperature
than for precipitation across all models.

\begin{figure}[t]
    \centerline{ \includegraphics[width=\textwidth]{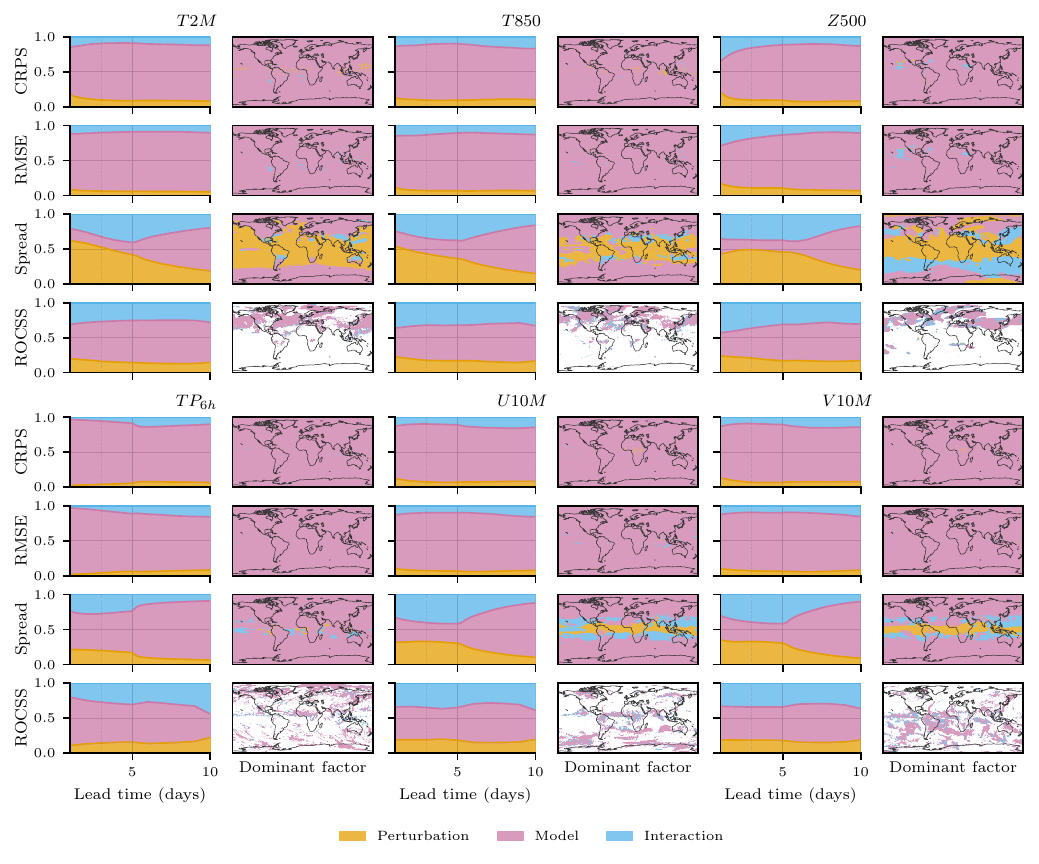}}
    \caption{Variance decomposition of forecast performance into
    contributions from perturbation method, model, and their
    interaction, using $\eta^{2}$ from a two-way ANOVA (see Appendix~\ref{metricsapp}). For
    each variable, results are shown for CRPS, RMSE, ensemble spread and ROCSS
    with left panels representing area-weighted global mean $\eta^{2}$ as a function
    of lead time and the right panel a spatial map of the dominant factor (perturbation,
    model, or interaction) at each grid point, determined by the largest
    $\eta^{2}$ component, averaged across lead times. For ROCSS we used the
    99th percentile, and the daily aggregates for each variable. Model
    architecture (pink) dominates CRPS and RMSE across nearly all variables
    and lead times, while the perturbation method (yellow) plays a larger role
    for ensemble spread, particularly at short lead times and for
    geopotential ($Z500$).}
    \label{perturbanalysis}
\end{figure}

In Fig.~\ref{perturbanalysis}, we investigate the relative role of the
perturbation and model choice from a statistical perspective, using analysis
of variance. For CRPS and RMSE, the model architecture is
overwhelmingly the leading factor and the corresponding spatial
maps are almost uniformly model-dominated. Perturbation methods contribute only
a thin marginal share (under 10\%), and the interaction term is similarly small. The picture changes
substantially for the ensemble spread, where the perturbation method explains a
large fraction of the variance at short lead times, between 30--50\%
for $T2M$, $T850$, $TP_{6h}$, $U10M$ and $V10M$ at 1 day lead time and then decays, 
so that by day~10, model choice again dominates. Geopotential
at 500\,hPa ($Z500$) is a partial exception: its spread is governed largely
by the model and perturbation interaction rather than by either
factor alone, with the interaction component visible across much of the global
map and particularly over the oceans. For ROCSS, model architecture remains the
leading factor across most variables (over 60\%), but the interaction term becomes more
prominent at longer lead times, and the spatial maps are sparser, reflecting the grid
points where the metric is undefined for rare events.

The choice of perturbation method exerts its strongest influence where expected: on the spread of the ensemble at short lead times, when initial-condition
uncertainty has not yet been overwhelmed by the model's error growth,
but it has comparatively little leverage on the accuracy of the ensemble
mean (RMSE) or the overall probabilistic skill (CRPS), both of which governed
almost entirely by the model choice. The case of $Z500$ spread, where the interaction
term dominates, suggests that some perturbation--model combinations behave non-additively,
which can be explained by the fact that HCBV and HENS use the forecast model itself
to generate the perturbed states, and $Z500$ is our seeding variable. 
The decay of the perturbation share with lead time is also consistent with a growing 
role of model-driven error at longer lead times.

For both case studies and global evaluations, the results reveal a consistent
pattern: any of the perturbations studied seem to improve the probabilistic
skill among deterministic AIWP models, with simpler methods (Gaussian and
Perlin) appearing to have a slight edge. Overall, ensemble skill (ROCSS, CRPS)
declines with forecast lead time, reflecting the inherent limits of predictability
and the growing influence of model-driven error. AIWP models perform
substantially better for temperature extremes (coherent, large-scale events)
than for precipitation extremes, which depend on small-scale
processes poorly represented in their training data (i.e., ERA5). Globally, CRPS and RMSE
metrics confirm that perturbation-based ensembles can emulate aspects of
probabilistic behavior but still lack the spatial variability and multiscale
energy structure of NWP ensembles or native probabilistic models, particularly for precipitation. 
Overall, while these perturbations move AIWP deterministic models toward meaningful ensemble
forecasting, performance mismatches persist.

\section{Discussion}
Our results are consistent with recent evaluations of deterministic AIWP models
\citep{olivettiDatadrivenModelsBeat2024, zhangNumericalModelsOutperform2025, brenowitzPracticalProbabilisticBenchmark2024},
which find that data-driven systems outperform traditional NWP on deterministic
metrics but show weaker probabilistic skill. As in
\citet{maheshHugeEnsemblesPart2025, maheshHugeEnsemblesPart2025a}, we confirm
that input perturbations can generate meaningful spread from AIWP architectures.
However, the comparable performance of HCBV and HENS to Gaussian and Perlin noise
contrasts with \citet{bulteUncertaintyQuantificationDatadriven2024}, who reported
Random Field Perturbations outperforming Gaussian noise; this discrepancy likely
reflects their use of a smaller amplification factor and spatially incoherent
noise. Our findings indicate that when perturbations are spatially coherent and
appropriately scaled, their specific structure matters less than their amplitude.
GraphCast shows an advantage for both precipitation and surface temperature,
suggesting benefits from graph-based atmospheric representations.

Ensemble power spectra (Appendix~\ref{powerspectra}) show that AIWP ensembles
underestimate precipitation spread at all scales relative to ENS, while
2m temperature spectra match more closely—particularly for FuXi–Perlin (Fig. \ref{powerspectrastdplot}).
Deterministic AIWP spectra track ERA5 well for temperature, but drop in power
at the mesoscale, and SFNO is underpowered across all scales for precipitation.
This is consistent with \citet{rodwellPowerSpectraPhysicsBased2025}.

Two caveats concern our choice of reference and baseline. First, ERA5 has known
limitations in capturing precipitation variability in the tropics and regions
with sparse observations \citep{laversEvaluationERA5Precipitation2022}, and the
2022 Pakistan event falls within both categories. Absolute ROCSS and CRPS values
may therefore be affected, though the relative ranking of models and perturbations
should be more robust since all configurations are evaluated against the same reference.
Gauge-based or satellite-derived products (e.g., \ IMERG) or regional reanalyses
with denser observational input would provide a more faithful evaluation
\citep{guptaMAUSAMObservationsfocusedAssessment2025}. 
Second, ECMWF's Open Data now provides high-quality ensemble initial conditions
that would likely improve performance and represent a stronger operational
baseline. We do not use them here because our focus is on self-contained
perturbation strategies applicable when operational ensemble analyses are
unavailable, such as in historical reforecasts; a direct comparison remains
for future work. 

The decline in ensemble skill with lead time reflects both initial-condition
sensitivity and model-induced error, which our perturbation-based ensembles do
not cleanly isolate: Gaussian and Perlin noise only crudely approximate the true
initial-condition distribution, while HCBV and HENS use the forecast model itself
to generate perturbed states, injecting model-dependent structure. Empirically,
the $\eta^{2}$ decomposition (Fig.~\ref{perturbanalysis}) shows that beyond
roughly five days, the perturbation method explains only a small fraction of
variance in CRPS and ensemble spread relative to model choice. This is consistent
with a growing role of model-dependent error at longer lead times (due to
autoregressive error accumulation), plausibly arising from architectural bias,
loss formulation, and underrepresented physical feedbacks. 
Combining input perturbations with model-checkpoint ensembles has been shown to
yield higher probabilistic skill \citep{maheshHugeEnsemblesPart2025,
bano-medinaCalibratedEnsemblesNeural2025}, but producing multiple checkpoints
or very large ensembles is computationally expensive. Alternative approaches to
introducing model stochasticity via latent-space, generative perturbations, or multi-model ensembles remain a promising research direction.

From a practical standpoint, the similarity in probabilistic skill between
state-independent (Gaussian, Perlin) and flow-dependent (HCBV, HENS)
perturbations is important considering the computational cost. Bred-vector
methods require additional forward passes ($d=3$ breeding cycles in our
configuration), whereas Gaussian and Perlin perturbations are essentially
free. Given that these simpler methods match or slightly exceed
flow-dependent approaches on global metrics, they offer the most favorable
cost trade-off for deterministic AIWP models, at least at $1^{\circ}$
resolution and 10-day horizons. Ensemble size is a second practical lever:
the decrease of importance of perturbation choice on spread and the
dominance of model choice beyond day 5 suggest that smaller ensembles may
suffice for short-range applications, with larger ensembles reserved for
tail discrimination \citep{maheshHugeEnsemblesPart2025,
bano-medinaCalibratedEnsemblesNeural2025}. An ensemble
size sensitivity analysis (Fig. \ref{skillcost}) revealed that AIFS ENS dominates
the skill--compute trade-off across all variables and ensemble sizes.
However, AIFS ENS requires 38\,GB of GPU memory by default \citep{EcmwfAifsens10Hugging2026}. On hardware with less memory the latency or run time will be higher. At short lead times, for 2m temperature, FuXi and GraphCast with Gaussian or Perlin perturbations approach (though do not reach) AIFS ENS skill with 30 ensemble members at a substantially lower memory footprint.
A systematic study of the skill--cost frontier would be valuable future
work. Input perturbations can be useful when native
probabilistic models are not available, or when memory or latency is the
binding constraint, especially at short lead times.
    
A broader pattern emerges from our results: spatially correlated noise, when
appropriately scaled, appears sufficient to generate skillful AIWP ensembles
regardless of its specific structure. Gaussian and Perlin perturbations differ
substantially in their spectral properties (Appendix~\ref{perturbations}) yet
yield spread and skill comparable to flow-dependent bred vectors, suggesting
that spatial coherence matters more than alignment with local flow. In \citet{bano-medinaCalibratedEnsemblesNeural2025}, they describe how white noise (spatially uncorrelated) perturbations eventually map onto the growing modes of the forecast, after some autoregressive steps. Our results seem to suggest that providing spatially correlated noise is sufficient to allow the projection to happen in the first forecast step. This is however
variable-dependent: FuXi and GraphCast ensembles match or exceed ENS on global
temperature CRPS up to 3--5 days, but precipitation skill degrades more
quickly. Predictive performance therefore depends not only on event rarity but
also on the physical nature of the target
\citep{zhaoExtensionWeatherBench22025}. Temperature extremes are often
spatially coherent and governed by persistent synoptic features
\citep{yangRelatingAnomalousLargescale2020}, which AIWP models capture well;
precipitation extremes, in contrast, depend on subgrid processes
underrepresented in training data \citep{hessDeepLearningImproving2022}.
Improving AIWP generalization for rare hydro-meteorological extremes may
therefore require targeted loss weighting, balanced sampling, or training with
precipitation-specific datasets
\citep{sunPredictingTrainingData2025, hessDeepLearningImproving2022}.

Model performance also depends on feature representation rather than the region
alone: FuXi captures the China heatwave well, but underperforms on the
monsoon-driven Pakistan precipitation. Evaluating AIWP ensembles on a feature
basis linked to atmospheric regimes or teleconnection patterns could yield more
diagnostic insight than regional averages. This study is further limited
in time scale, geography, and resolution; extensions to seasonal variability,
other regions, and higher resolutions (given that $1^{\circ}$ may penalize
precipitation performance) are natural next steps
\citep{hessDeepLearningImproving2022}.

Overall, global metrics and case studies give complementary views: globally,
perturbation-based AIWP ensembles are broadly competitive on average, while
case studies expose where that skill translates into useful discrimination of
high-impact events. Deterministic AI weather models can approximate
aspects of probabilistic behavior through input perturbations such as simple
spatially coherent noise and bred-vector approaches, but still fall short of
fully capturing uncertainty and extremes. Gaussian and Perlin methods offer a
promising bridge toward more realistic ensemble forecasts at low computational
cost. Future work could integrate spatially coherent perturbations with
generative or latent-space stochastic approaches, improve training data for extremes, or explore multi-model ensembles.

\section{Conclusions}
This study evaluated how deterministic AIWP models respond to input
perturbations and how their resulting ensembles represent uncertainty
during extreme events. Using FuXi, GraphCast, and SFNO, we generated
50-member ensembles with Gaussian, Perlin, HCBV, and HENS
perturbations, evaluated globally over August 2022 and on the 2022
Pakistan floods and China heatwave case studies. To our knowledge,
this is the first multi-model, multi-perturbation comparison of
ensemble generation strategies for deterministic AIWP models that
combines case-study and global threshold analyses.

Our central finding is that the dominant factor governing ensemble
skill is the underlying model architecture, while the choice of
perturbation method plays a secondary role: simpler methods produce
similarly skillful ensembles to flow-dependent ones when appropriately
scaled. Perturbation-based ensembles narrow but do not close the gap
to NWP (ENS) or to native probabilistic AIWP models such as AIFS ENS,
which remains the recommended choice where hardware permits.
Deterministic AIWP models with simple perturbations are most
attractive when memory or latency is a constraint. Across variables, all AIWP
models captured temperature extremes more reliably than precipitation,
and probabilistic skill declined with lead time.

Overall, perturbation-based ensembles extend deterministic AI models
toward operational probabilistic forecasting. However, fully capturing
the uncertainty and multiscale dynamics of extreme events could
require hybrid strategies that integrate input perturbations with
latent-space perturbations and generative approaches, improved
training on rare extremes, or multi-model ensemble frameworks.
Advancing along these directions will be key for building trustworthy,
AI-driven early warning systems for high-impact weather events.

\section*{Acknowledgments}
This research has been supported by the European Union’s Horizon
Europe research and innovation program (EU Horizon Europe) project MedEWSa under
grant agreement no. 101121192. M.-Á. F.-T. thanks the Universidad Carlos III de Madrid, Spain, for the support provided through the Funding for Research Activities Program (project 2025/00803/001) and the Regional Government of Madrid through project TEC-2024/COM-322, and also acknowledges the computer resources provided by Artemisa, funded by the ERDF and Comunitat Valenciana, as well as the technical support provided by the Instituto de Física Corpuscular, IFIC (CSIC-UV).

%
%
\section*{Data availability statement}
The ERA5 data in this study was obtained using the ERA5 ARCO dataset hosted on Google Cloud \texttt{https://github.com/google-research/arco-era5}. The IFS ENS data in this study was obtained from the WeatherBench 2 dataset \texttt{https://weatherbench2.readthedocs.io/en/latest/data-guide.html}. The code for processing the data, running the perturbations, and computing the metrics can be found in \texttt{https://gitlab.hhi.fraunhofer.de/ai-aml/aiwpuq}.

\clearpage

\appendix
\section{Evaluation Metrics, Statistical Significance Testing and Variance Decomposition}
\applabel{A}{metricsapp}

    For the following metric definitions, $y_{nt}$ is the ERA5 value for each
    spatial location $n$ ($N$ spatial locations in a study region) at time
    $t \in \{1, \dots , T\}$, with $T$ as the number of forecast timesteps, and
    $\hat{y}_{nt}^{(m)}$ as the forecast provided by an ensemble member
    $m \in \{1, \dots, M\}$, where $M=50$.

    \subsection{Root Mean Squared Error (RMSE)}
    The Root Mean Squared Error (RMSE) is a metric that measures the average magnitude
    of errors between predicted and observed values \citep{raspWeatherBench2Benchmark2024a}.
    It penalizes larger errors more heavily, making it sensitive to outliers. Lower
    RMSE represents a better forecast. RMSE is defined as follows:

    \begin{equation}
        \text{RMSE}= \sqrt{ \frac{1}{NT}\sum_{n=1}^{N}\sum_{t=1}^{T}\left( y_{nt}-
        \frac{1}{M}\sum_{m=1}^{M}\hat{y}_{nt}^{(m)}\right)^{2}}
    \end{equation}

    \subsection{Continuous Ranked Probability Score (CRPS)}
    The Continuous Ranked Probability Score (CRPS) is a scoring rule that measures
    the accuracy of probabilistic forecasts by quantifying the difference between
    the predicted cumulative distribution function (CDF) and the observed outcome
    \citep{raspWeatherBench2Benchmark2024a}. It generalizes the mean absolute error
    to probabilistic forecasts. Lower CRPS indicates better forecast skill. CRPS
    is defined as follows:

    \begin{equation}
        \text{CRPS}= \frac{1}{NT}\sum_{n=1}^{N}\sum_{t=1}^{T}\left[ \frac{1}{M}\sum
        _{m=1}^{M}\left| \hat{y}_{nt}^{(m)}- y_{nt}\right| - \frac{1}{2M^{2}}\sum
        _{m=1}^{M}\sum_{k=1}^{M}\left| \hat{y}_{nt}^{(m)}- \hat{y}_{nt}^{(k)}\right
        | \right]
    \end{equation}

    \subsection{Receiver Operating Characteristic Skill Score (ROCSS)}
    The Receiver Operating Characteristic Skill Score (ROCSS) is a metric
    derived from the Area Under the ROC Curve (AUC) that quantifies the ability
    of a forecasting system to discriminate between the occurrence and non-occurrence
    of an event, relative to random chance \citep{zhaoExtensionWeatherBench22025}.

    We set the event threshold $\tau$ to the 99th climatological percentile from
    ERA5 (1990–2020) to capture the extremes relevant to this study. The 2022 Pakistan
    floods were a record-breaking precipitation event \citep{Singh2024-wk}, warranting
    a high percentile threshold to isolate tail behavior. For heat extremes in
    China, the 99th percentile aligns with established heatwave-impact studies such
    as \citet{sunHeatWaveCharacteristics2021}, where severe heatwave conditions are
    characterized using upper-percentile exceedance. Together, these justify the
    choice of $\tau=P_{99}$ as an appropriate discriminator for rare, high-impact
    extremes.

    The computation of the ROCSS proceeds as follows:

    \begin{enumerate}
        \item Thresholding observations and ensemble forecasts: Given the
            threshold $\tau$, observations and ensemble forecasts are converted into
            binary outcomes. Then, ensemble binary forecasts are converted into forecast
            probabilities:

            \begin{equation}
                o_{nt}= \mathbf{1}\{ y_{nt}> \tau \}, \quad f_{nt}^{(m)}= \mathbf{1}
                \{ \hat{y}_{nt}^{(m)}> \tau \}, \quad r_{nt}= \sum_{m=1}^{M}f_{nt}
                ^{(m)}, \quad p_{nt}= \frac{r_{nt}}{M+1}
            \end{equation}

            Here, $o_{nt}$ is the binary observation, $f_{nt}^{(m)}$ is the
            binary forecast of ensemble member $m$, $r_{nt}$ is the ensemble “vote
            count” and $p_{nt}$ is the resulting forecast probability using Weibull’s
            plotting position \citep{weibull1939astatistical}.

        \item Applying probability thresholds: To construct the ROC curve, the forecast
            probability is converted back to binary outcomes for each probability
            threshold $\theta \in [0,1]$:
            \begin{equation}
                \hat{o}_{nt}(\theta) = \mathbf{1}\{ p_{nt}\geq \theta \}
            \end{equation}

        \item Calculating Hit Rate (HR) and False Alarm Rate (FAR):
            \begin{equation}
                \mathrm{HR}(\theta)=\frac{\sum_{n,t}\mathbf{1}\{\hat{o}_{nt}(\theta)=1\land
                o_{nt}=1\}}{\sum_{n,t}o_{nt}},\qquad \mathrm{FAR}(\theta)=\frac{\sum_{n,t}\mathbf{1}\{\hat{o}_{nt}(\theta)=1\land
                o_{nt}=0\}}{\sum_{n,t}(1-o_{nt})}
            \end{equation}

        \item Constructing the ROC curve: The ROC curve plots the trade-off
            between hits and false alarms across all probability thresholds:
            \begin{equation}
                \{ (\text{FAR}(\theta), \text{HR}(\theta)) : \theta \in [0,1] \}
            \end{equation}

        \item Computing the AUC:
            \begin{equation}
                AUC_{\text{forecast}}= \int_{0}^{1}\text{HR}(\text{FAR}^{-1}(x))
                \, dx
            \end{equation}
            \noindent
            where $\text{FAR}^{-1}(x)$ denotes the inverse mapping of the false
            alarm rate along the ROC curve.

        \item Computing the ROCSS: Finally, the ROCSS compares the forecast AUC to
            a reference (random) forecast, which has
            $AUC_{\text{reference}}= 0.5$:
            \begin{equation}
                \text{ROCSS}= \frac{AUC_{\text{forecast}}- AUC_{\text{reference}}}{1
                - AUC_{\text{reference}}}
            \end{equation}

            A higher ROCSS indicates better discriminatory skill, with 1 representing
            perfect skill and 0 corresponding to no skill.
    \end{enumerate}

    \subsection{Statistical Significance Testing}

    To assess whether the differences in model performance between AIWP model
    ensembles and ENS, specifically in ROCSS, are statistically significant, we employ
    a clustered panel \textit{t}-test based on a panel regression framework, following
    \citet{olivettiDatadrivenModelsBeat2024}.

    For each variable (daily accumulated precipitation, $TP_{24h}$, and maximum daily
    temperature, $T2M_{24h}$) and forecast lead time, we first compute the
    difference in ROCSS between the AIWP model ensemble and ENS at each grid
    cell and validation date. The resulting dataset ($\text{diff}$) is structured
    as a panel with spatial and temporal dimensions, indexed by latitude and
    longitude coordinates and valid time:

    \begin{equation}
        \text{diff}_{nt}= y_{nt}
    \end{equation}

    \noindent
    where $n$ denotes the spatial location (grid point) and $t$ the time index (validation
    date). We then estimate the following panel model:

    \begin{equation}
        y_{nt}= \alpha + \varepsilon_{nt}
    \end{equation}

    \noindent
    where $\alpha$ represents the mean difference in ROCSS between the AIWP model
    ensemble and ENS, and $\varepsilon_{nt}$ is the error term. The model is fit
    with two-way clustered standard errors, accounting for correlation within
    both spatial entities (grid cells) and across time. From the fitted model,
    we extract the \textit{t}-statistic associated with the intercept ($\alpha$),
    representing whether the mean difference significantly differs from zero.

    Significance thresholds are determined from the Student’s \textit{t} distribution
    with degrees of freedom equal to the number of panel observations.
    Differences are classified as statistically significant at the 5\% level ($p
    < 0.05$) when the \textit{t}-statistic exceeds these bounds. The \texttt{PanelOLS}
    implementation from the \texttt{linearmodels} Python package is used to perform
    the computations.

    \subsection{Variance Decomposition}

    To quantify the relative importance of perturbation method and model
    architecture in determining forecast performance, we perform a two-way analysis
    of variance (ANOVA) at each grid point and lead time. The two factors are
    the perturbation method $P$ (Gaussian, Perlin, HCBV, HENS; $p = 4$ levels)
    and the model architecture $M$ (FuXi, GraphCast, SFNO; $m = 3$ levels). For
    each combination of factors, the response variable $Y_{ij}$ represents the value
    of a given metric (CRPS, RMSE, or ensemble spread) for perturbation $i$ and model
    $j$. The total variability, or total sum of squares ($SS$) is decomposed into
    contributions from each factor:
    \begin{equation}
        SS_{\text{total}}= SS_{\text{perturbation}}+ SS_{\text{model}}+ SS_{\text{residual}}
    \end{equation}
    where:
    \begin{equation}
        SS_{\text{perturbation}}= m \sum_{i=1}^{p}(\bar{Y}_{i \cdot}- \bar{Y}_{\cdot
        \cdot})^{2}, \qquad SS_{\text{model}}= p \sum_{j=1}^{m}(\bar{Y}_{\cdot j}
        - \bar{Y}_{\cdot \cdot})^{2}
    \end{equation}
    \begin{equation}
        \\
        SS_{\text{residual}}= SS_{\text{total}}- SS_{\text{perturbation}}- SS_{\text{model}}
    \end{equation}
    with $\bar{Y}_{i \cdot}$ the mean over all models for perturbation $i$,
    $\bar{Y}_{\cdot j}$ the mean over all perturbations for model $j$, and $\bar{Y}
    _{\cdot \cdot}$ the grand mean. The residual term captures the interaction
    between the perturbation method and model architecture, reflecting non-additive effects
    where certain perturbation--model combinations perform differently than
    expected from their individual contributions. Because the design has no replicates, the interaction effect and the residual error are not separately identifiable, so any unmodeled noise is absorbed into the interaction term. 

    The effect size for each component is quantified using $\eta^{2}$, defined as
    the fraction of total variance explained:
    \begin{equation}
        \eta^{2}_{\text{perturbation}}= \frac{SS_{\text{perturbation}}}{SS_{\text{total}}}
        , \qquad \eta^{2}_{\text{model}}= \frac{SS_{\text{model}}}{SS_{\text{total}}}
        , \qquad \eta^{2}_{\text{interaction}}= \frac{SS_{\text{residual}}}{SS_{\text{total}}}
        .
    \end{equation}
    By construction, $\eta^{2}_{\text{perturbation}}+ \eta^{2}_{\text{model}}+ \eta
    ^{2}_{\text{interaction}}= 1$. A value of $\eta^{2}\approx 1$ for a given
    component indicates that it explains nearly all of the variance in the
    metric, while $\eta^{2}\approx 0$ indicates negligible influence. The
    analysis is performed independently at each grid point and lead time. The stacked
    area plots in Fig.~\ref{perturbanalysis} show the area-weighted global mean
    of $\eta^{2}_{\text{perturbation}}$, $\eta^{2}_{\text{model}}$, and
    $\eta^{2}_{\text{interaction}}$ as a function of lead time. The spatial maps
    display the dominant factor at each grid point, defined as the component
    with the largest $\eta^{2}$, averaged across all lead times.
    
    \clearpage

    \section{Additional plots}
    \applabel{B}{globalplots}

    \begin{figure}[h]
        \centerline{ \includegraphics[width=0.9\linewidth]{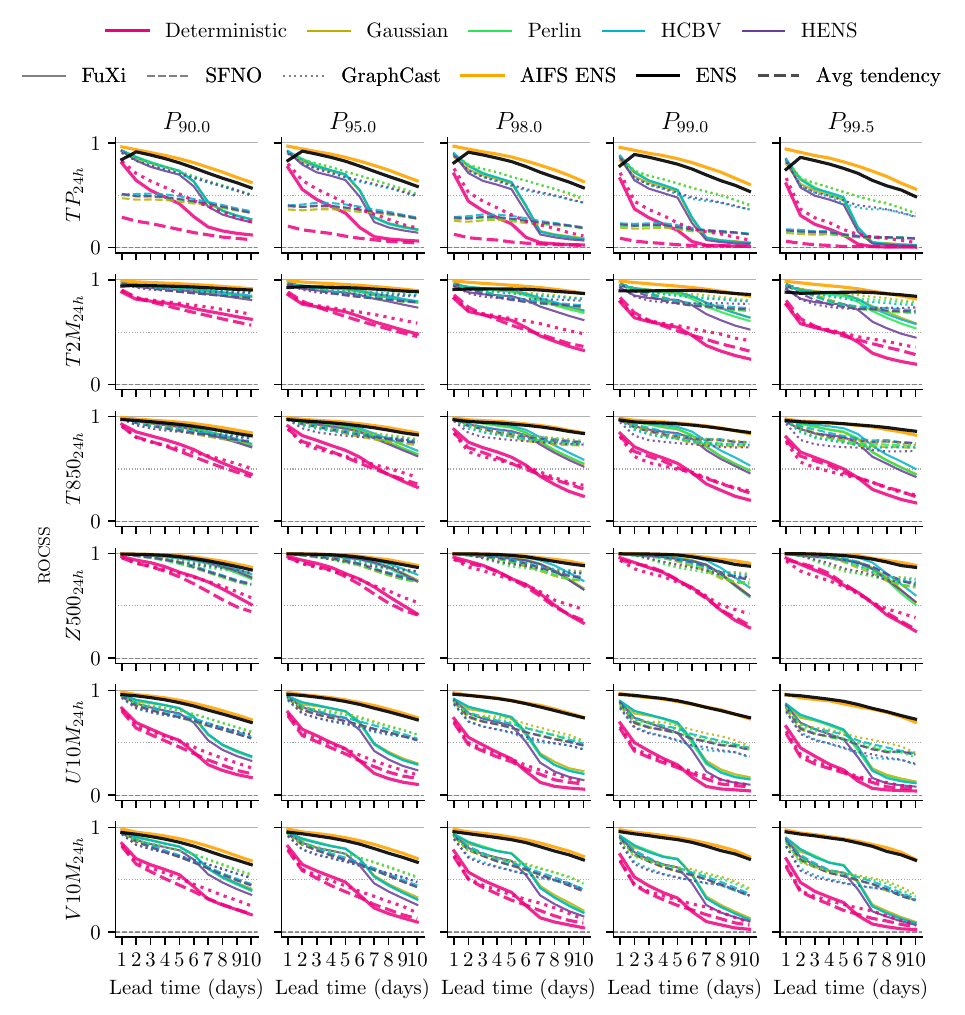} }
        \caption{ROCSS values at the 90th till the 99.5th percentile of the ERA5
        1990-2020 climatology for daily accumulated precipitation ($TP_{24h}$),
        maximum daily surface temperature ($T2M_{24h}$), temperature at 850hPa
        ($T850_{24h}$), geopotential at 500hPa ($Z500_{24h}$), and U and V 10 m
        height wind components ($U10M_{24h}$, $V10M_{24h}$), across 10 lead times,
        for the different AIWP models and their ensembles, globally. The higher the
        ROCSS values, the better the performance on the extremes.}
        \label{globalrocss}
    \end{figure}

    \begin{figure}[t]
        \centerline{ \includegraphics[width=0.8\textwidth]{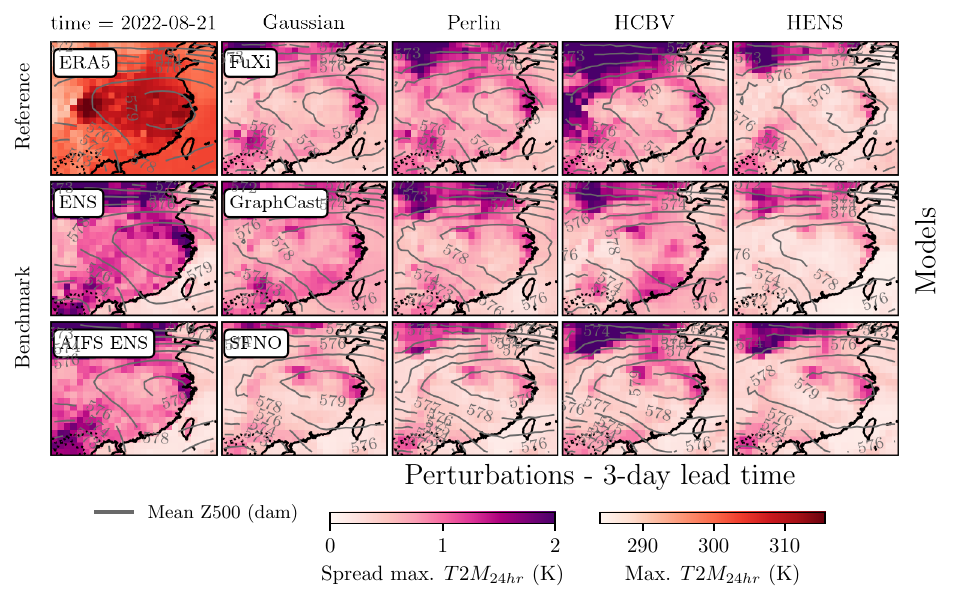} }
        \caption{Daily maximum temperature spread for the different ensemble
        models (ENS, AIWPs) over China on 21st August 2022 for a 3-day lead time
        forecast. ERA5 spatial distribution for maximum temperature
        is shown in the top left corner. Daily average geopotential height at 500
        hPa is also displayed as contour lines. AIFS ENS closely matches the ENS
        spread, while FuXi–Perlin comes closest among the perturbed ensembles.}
        \label{spreadchina}
    \end{figure}

    \begin{figure}[h]
        \centerline{\includegraphics{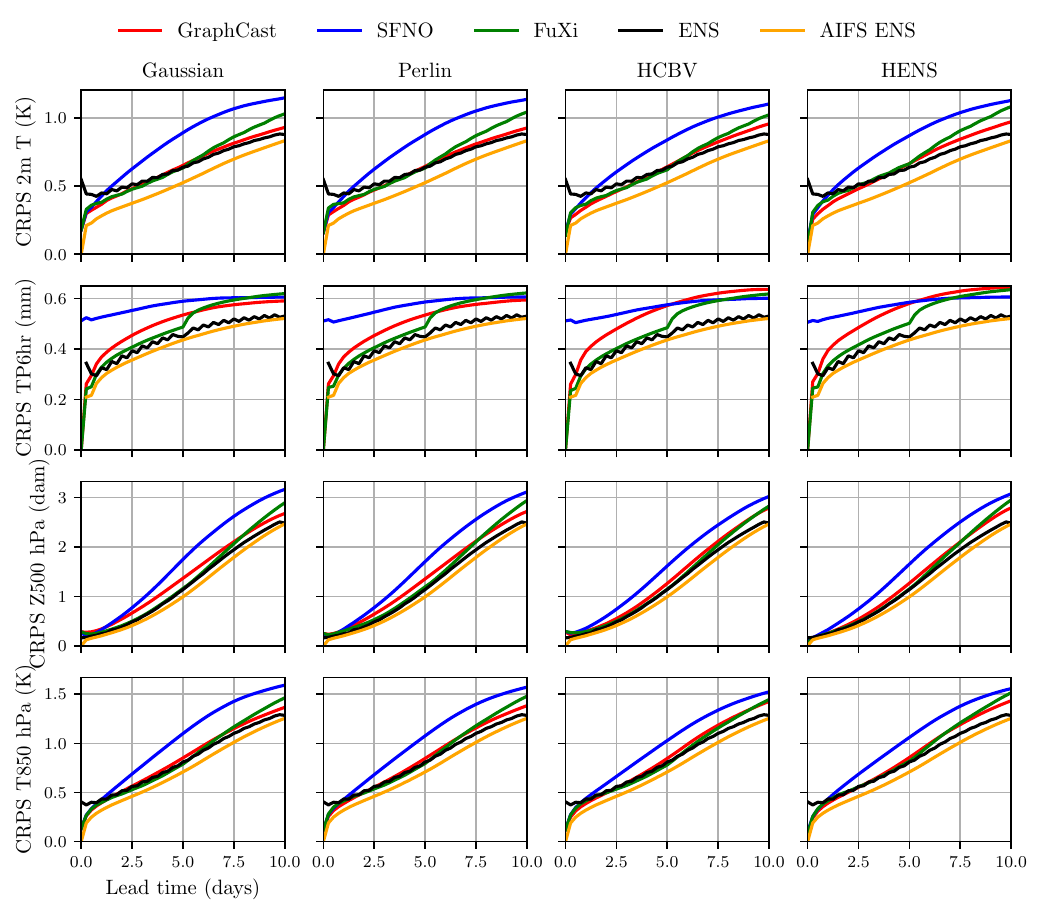} }
        \caption{Global average CRPS over 10-day lead time for the different
        perturbation methods and AIWP models and ENS, in August 2022.}
        \label{crpsplot}
    \end{figure}

    \begin{figure}[h]
        \centerline{\includegraphics{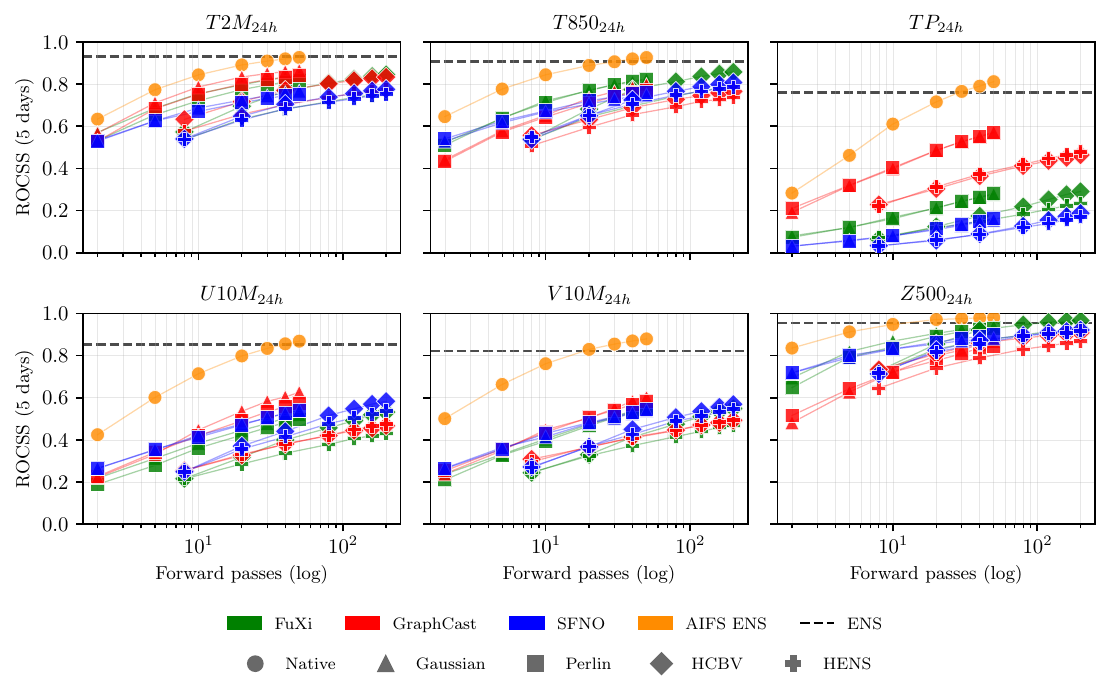} }
        \caption{ROCSS for $P_{99}$ and number of forward passes at 5 days lead time, for all six evaluation variables. Each marker corresponds to one (model, perturbation, ensemble size) configuration with color encoding the model and shape the perturbation method. Ensemble size is $N \in \{2, 5, 10, 20, 30, 40, 50\}$. For $N < 50$, ROCSS is computed on a single random sample of $N$ members drawn from the full 50-member ensemble of each configuration. The $x$-axis reports the number of forward passes $C$ required to produce the forecast, used as a model and hardware independent compute cost approximation: $C = N$ for
      Native (AIFS ENS), Gaussian, and Perlin perturbations, and $C = N\,(1+d)$ with
      $d = 3$ breeding cycles per member for HCBV and HENS. The
      dashed line is the IFS ENS reference ROCSS. AIFS ENS dominates the skill--compute trade-off
      across all variables and ensemble sizes. For temperature at low budgets ($N=30$), FuXi and GraphCast with Gaussian or Perlin are competitive with HCBV and HENS, and perform close to AIFS ENS, supporting the use of simple input perturbations when AIFS ENS is not available.}
        \label{skillcost}
    \end{figure}

    \clearpage

    \section{Power Spectra of Forecasts}
    \applabel{C}{powerspectra}

    \begin{figure}[h]
        \centerline{ \includegraphics{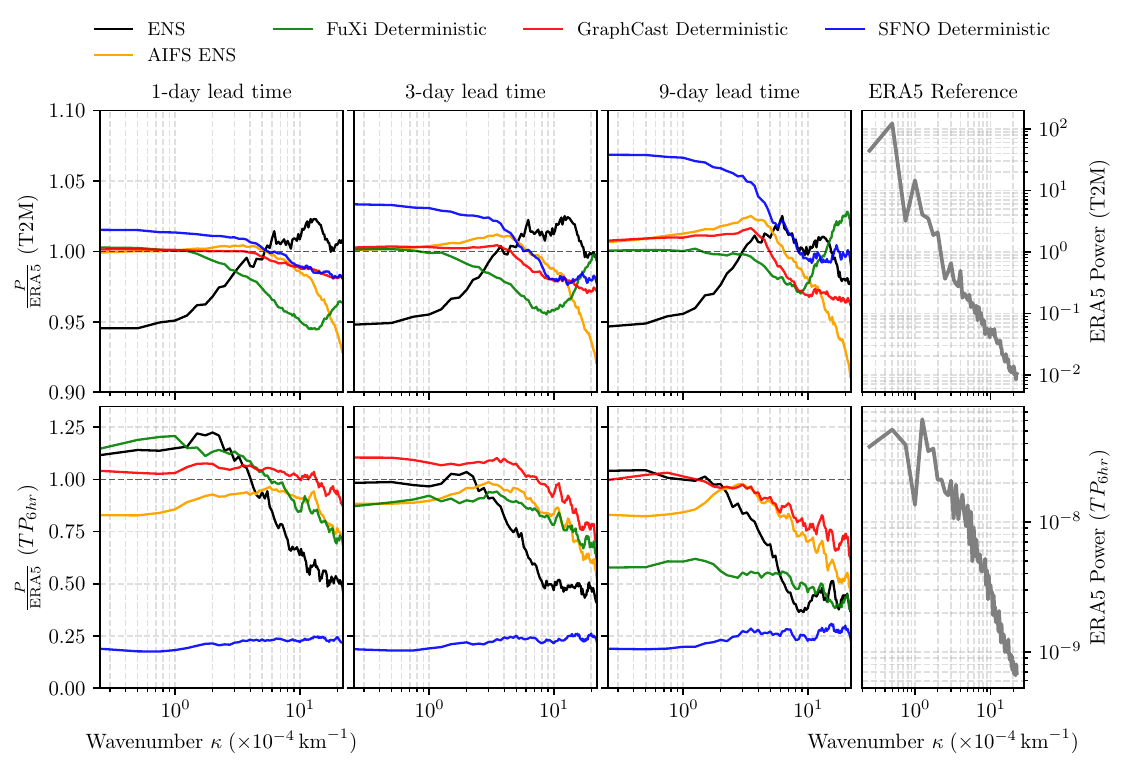} }
        \caption{Spectral comparison of AIWP models against the ERA5 reanalysis
        for 2~m temperature (T2M, top row) and 6-hour accumulated precipitation
        ($TP_{6h}$, bottom row). Each panel shows the smoothed power ratio, $\frac{P}{\mathrm{ERA5}}$,
        as a function of wavenumber $\kappa$ ($\times10^{-4}\,\mathrm{km}^{-1}$),
        for the ensemble mean of ENS and AIFS ENS and the deterministic forecasts for FuXi,
        GraphCast, and SFNO, at forecast lead times of 1, 3, and 9~days. Values above
        1 indicate excess variance relative to ERA5 (over-energetic spectra),
        while values under 1 indicate reduced small-scale variability (over-smoothing).
        The rightmost column displays the reference ERA5 power spectra for each variable.
        }
        \label{powerspectraplot}
    \end{figure}

    \begin{figure}[h]
        \centerline{ \includegraphics{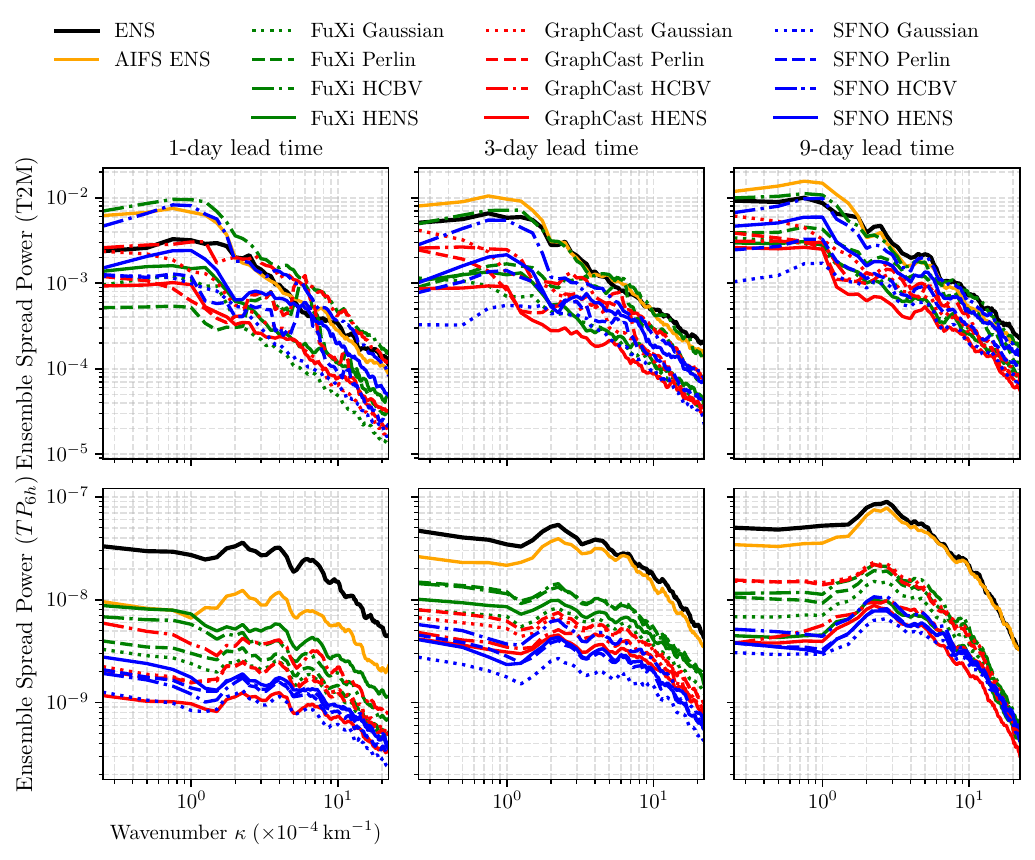} }
        \caption{Ensemble spread spectral comparison for 2~m temperature (T2M, top
        row) and 6-hour accumulated precipitation ($TP_{6h}$, bottom row). Each
        panel shows the smoothed ensemble spread spectral power, as a function of
        wavenumber $\kappa$ ($\times10^{-4}\,\mathrm{km}^{-1}$), for ENS and AIFS ENS,
        and the perturbed forecasts for FuXi, GraphCast, and SFNO, at forecast lead
        times of 1, 3, and 9~days, for a single initialization time in August
        15th, 2022.}
        \label{powerspectrastdplot}
    \end{figure}

    \clearpage
    
    \section{Visualization of perturbations}
    \applabel{D}{perturbations}

    \begin{figure}[h]
        \centerline{\includegraphics[width=0.8\linewidth]{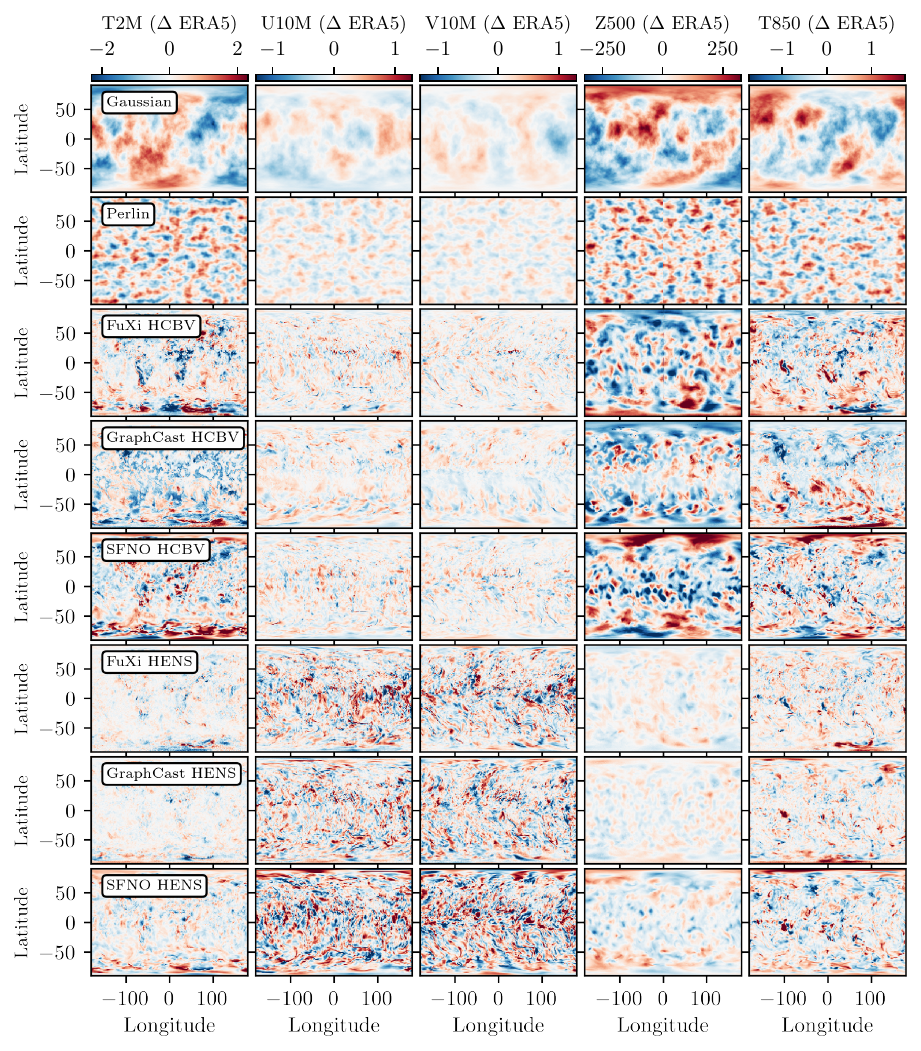}}
        \caption{Visualization of the initial-condition perturbations applied to
        each AIWP model. Rows correspond to the different perturbation strategies:
        state-independent Gaussian and Perlin noise (shared across models), and the
        flow-dependent bred-vector methods HCBV and HENS, shown separately for
        FuXi, GraphCast, and SFNO. Columns correspond to the perturbed input variables,
        each displayed as a difference with respect to the ERA5 initial
        condition ($\Delta$ERA5), with red (blue) indicating positive (negative)
        values.}
        \label{perturbs}
    \end{figure}

    \clearpage

    \section{Sensitivity analysis of the perturbation scaling factor}
    \applabel{E}{scaling}

    All four perturbation methods used in this study include a dimensionless scaling
    factor $s$ that controls the overall magnitude of the perturbation applied to
    the initial state (Section~3\ref{perturbation}). To select an appropriate value of $s$ for
    each method, we performed a sensitivity analysis, described below.

\begin{figure}[h]
        \centerline{\includegraphics{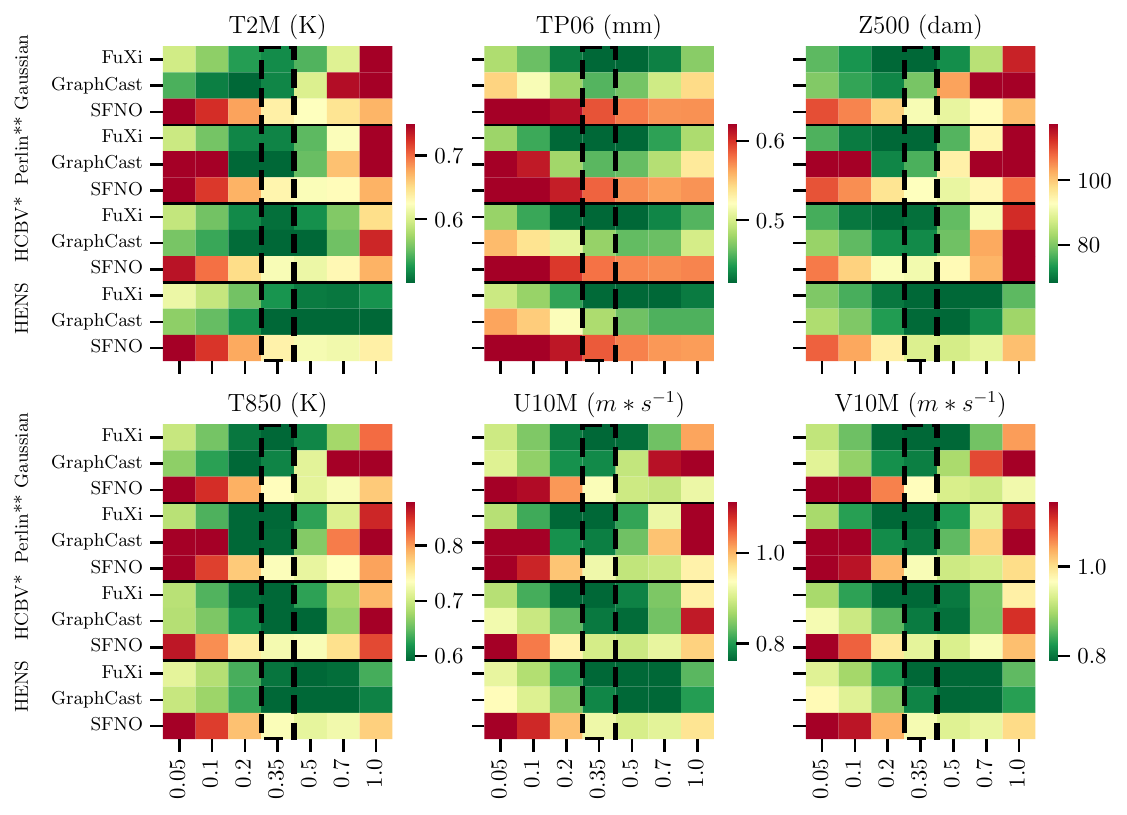}}
        \caption{Mean CRPS across lead times as a function of the perturbation
        scaling factor $s$, for each combination of variable ($T2M$, $TP_{6h}$,
        $Z500$, $T850$, $U10M$, $V10M$), perturbation method (Gaussian, Perlin,
        HCBV, HENS), and the AIWP model (FuXi, GraphCast, SFNO). Lower CRPS
        indicates better probabilistic skill. For HCBV, the effective scaling
        factor is one order of magnitude smaller than the value shown on the
        axis (e.g., an axis value of $0.35$ corresponds to an applied factor of $0
        .035$); for Perlin, the effective scaling factor is one quarter of the
        axis value (e.g., $0.35$ corresponds to $0.0875$). The vertical dashed line
        marks the selected axis value of $s = 0.35$, which minimizes CRPS across
        most variable--model combinations and was used for all results reported in
        the main text.}
        \label{scalingheatmap}
    \end{figure}

\begin{figure}[h]
        \centerline{\includegraphics{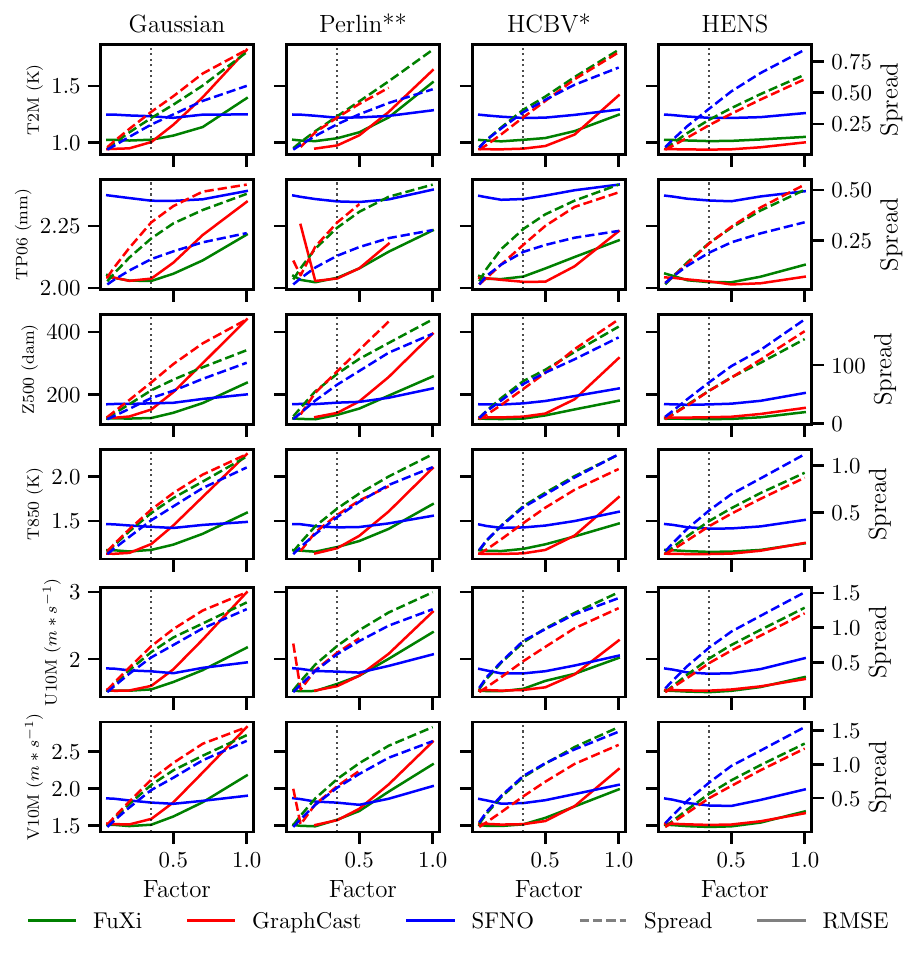}}
        \caption{Ensemble standard deviation (spread, dashed) and ensemble-mean RMSE
        (solid) as a function of the perturbation scaling factor $s$, for each
        variable and perturbation method, with one curve per AIWP model (FuXi, GraphCast,
        SFNO). For HCBV, the effective scaling factor is one order of magnitude smaller
        than the value shown on the axis; for Perlin, it is one quarter of the axis
        value. The vertical dotted line marks the selected axis value of
        $s = 0.35$, which maximizes spread without increasing RMSE,
        across variable--model combinations.}
        \label{scalingspreadrmse}
    \end{figure}

    We used a subset of our evaluation set for the scaling analysis: four initialization times on 15~August~2022 (00, 06, 12
    and 18~UTC), 10 ensemble members per configuration, and a 5-day forecast rollout.
    The sweep was repeated for all three AIWP architectures (FuXi, GraphCast, SFNO)
    and all four perturbation methods (Gaussian, Perlin, HCBV, HENS). For Gaussian and HENS perturbations, $s$ was
    varied directly over the values shown in Figs.~\ref{scalingheatmap} and~\ref{scalingspreadrmse}.
    For HCBV, the effective scaling is one order of magnitude smaller than the
    value plotted on the axis (e.g., an axis value of $0.35$ corresponds to an applied
    factor of $0.035$); for Perlin, the effective scaling is one quarter of the axis
    value (e.g., $0.35$ corresponds to $0.0875$).

    For each combination of perturbation method, model, and variable, we computed
    (i) the mean CRPS across lead times, (ii) the ensemble mean RMSE, and (iii)
    the ensemble standard deviation (spread). A good scaling factor should
    minimize CRPS and RMSE while producing a spread comparable to the ensemble error,
    in line with standard spread-skill consistency requirements. Across variables
    and models, we found that an axis value of $s = 0.35$ best satisfies these criteria
    for all four methods (corresponding to applied factors of $0.35$ for
    Gaussian and HENS, $0.035$ for HCBV, and $0.0875$ for Perlin). This value is
    also broadly consistent with the scaling adopted by \citet{maheshHugeEnsemblesPart2025}
    for HENS. All results reported use these settings.

    \clearpage

    \bibliographystyle{unsrtnat}
    \bibliography{updated}

\end{document}